\documentclass[11pt,a4paper]{article}
\usepackage[margin=26mm]{geometry}
\usepackage{amsmath,amssymb,amsthm,mathtools,mathrsfs}
\usepackage{xcolor,booktabs,enumitem}
\usepackage[colorlinks=true,linkcolor=blue!55!black,citecolor=blue!55!black,urlcolor=blue!55!black]{hyperref}
\usepackage{microtype}
\newtheorem{prop}{Proposition}[section]
\newtheorem{corollary}[prop]{Corollary}
\newtheorem{lemma}[prop]{Lemma}
\newtheorem{theorem}[prop]{Theorem}
\newtheorem{maintheorem}{Theorem}

\theoremstyle{definition}
\newtheorem{definition}[prop]{Definition}
\theoremstyle{remark}

\newtheorem{remark}[prop]{Remark}
\newtheorem{example}[prop]{Example}
\newcommand{\MC}{\operatorname{MC}}
\newcommand{\PU}{\operatorname{PU}}
\newcommand{\SU}{\operatorname{SU}}
\newcommand{\tr}{\operatorname{tr}}
\newcommand{\Q}{\mathbb Q}
\newcommand{\C}{\mathbb C}

\newcommand{\cO}{\mathcal O}
\newcommand{\LL}{L}
\newcommand{\RR}{R}
\newcommand{\len}{\ell}
\newcommand{\qout}{q_{\mathrm{out}}}
\title{Universal Complex Reflection Gates from Middle Convolution:\\Braid Dynamics and Exact Synthesis Costs}
\author{Haru Negami\\\small Japan Agency for Marine-Earth Science and Technology\\\small (JAMSTEC)}
\date{26 November 2026}
\begin{document}
\makeatletter
\renewcommand{\@maketitle}{\begin{center}
{\LARGE \@title\par}\medskip
{\large \begin{tabular}{c}\@author\end{tabular}\par}
\smallskip \@date
\end{center}}
\makeatother
\maketitle
\begin{abstract}
We construct quantum gates that act as complex reflections. Each multiplies a unit vector by a nontrivial phase and fixes its orthogonal complement pointwise. We obtain these gates by multiplicative middle convolution from two-dimensional free-group representations generated by reflections. We prove three main results. (A) We parametrize the braid orbit of the initial trace triple and determine the degrees of the resulting trace polynomials. As a consequence, we recover the stretch factor of any pseudo-Anosov three-braid from the valuations of a single trace coordinate after the first two iterates of a fixed initial triple. (B) Under an explicit positivity condition, the gates form a universal gate set. For rational overlaps between input normals, braid dynamics give lower bounds on shortest word lengths in the initial generators and their inverses, up to global phase. Under iteration of a pseudo-Anosov three-braid, the sum of the shortest lengths of any two of the three evolving gates grows at an exponential rate equal to the stretch factor of the braid. (C) In rank three, a faithful free-product representation gives exact Fibonacci formulas for the individual shortest lengths under $\sigma_1\sigma_2^{-1}$ at all iterates for transcendental overlaps and up to any prescribed iterate for rational overlaps with sufficiently large denominators.
\par\smallskip
\noindent\textit{2020 Mathematics Subject Classification.} Primary 20F36. Secondary 37P05, 81P65.\par
\noindent\textit{Keywords.} Braid group actions, middle convolution, complex reflections, trace dynamics, exact quantum gate synthesis.
\end{abstract}

\section{Introduction and main results}
A quantum gate is a unitary operator on a state space $\mathcal H$, for example $\mathcal H=(\mathbb C^2)^{\otimes m}$ for $m$ qubits. A unitary representation $\rho:G\to U(\mathcal H)$ assigns a gate to each group element. A word in the generators of $G$ gives a product of their matrices and inverses. What is the shortest word that represents the same gate?

We count each generator in a fixed gate set and its inverse as one letter and identify gates up to global phase. The minimum length is the exact synthesis cost. For two targets, the sum of their individual costs is the pair cost. A universal gate set can approximate every unitary on $\mathcal H$ up to phase. We consider two groups in particular.

\begin{definition}[Free groups and braid groups]\label{def:freebraid}
The free group $F_n$ consists of reduced words in $x_1^{\pm1},\ldots,x_n^{\pm1}$, with multiplication by concatenation and reduction.

\noindent The braid group $B_n$ on $n$ strands has generators $\sigma_1,\ldots,\sigma_{n-1}$ and defining relations
\[
\begin{aligned}
  \sigma_i\sigma_j             &= \sigma_j\sigma_i             && (|i-j|\ge 2),\\
  \sigma_i\sigma_{i+1}\sigma_i &= \sigma_{i+1}\sigma_i\sigma_{i+1} && (1\le i\le n-2).
\end{aligned}
\]
\end{definition}

In topological quantum computing, exchanging anyons implements gates through unitary braid group representations \cite{FLW}. Here the gates themselves are images of free generators, so the construction also applies to free-group representations without a compatible braid action. A braid $\beta$ acts on the free group by an automorphism $\alpha_\beta$. Evaluating the iterated word $\alpha_\beta^k(x_j)$ gives the target gate $\rho(\alpha_\beta^k(x_j))$ as an explicit product of the initial gates and their inverses.

Consider the three gates $A,B,C$ of our cyclotomic family on a qutrit, a three-level system with a chosen computational basis. Fix an odd $N\ge7$. Write $s_0=(a,a,b)$ for the initial triple of normalized pair traces in the order $12,23,13$. On gate triples the moves act by $L(A,B,C)=(ABA^{-1},A,C)$ and $R(A,B,C)=(A,C,C^{-1}BC)$, where $L=\sigma_1$ and $R=\sigma_2^{-1}$ in $B_3$. Both $R$ and $LR$ (applying $R$ first) send $s_0$ to $(b,a,a)$. 

\noindent\begin{minipage}{\linewidth}
Let $0<c<1$ be the overlap of the two distinct input reflection normals. For transcendental $c$, the first and third targets have the minimum lengths shown below (Theorem~\ref{thm:sharpwords}(ii), $n=1$).
\[
\begin{array}{c|cc|c}
 \text{braid} & \text{first target} & \text{third target} & \text{sum of minimum lengths}\\ \hline
 R  & A         & C^{-1}BC & 1+3=4\\
 LR & ACA^{-1}  & C^{-1}BC & 3+3=6
\end{array}
\]
\end{minipage}\par

The same minima hold at $N=13$, $c=3/5$ by reduction modulo $53$ (Proposition~\ref{prop:f53}). Thus the three pair traces alone do not determine the minimum lengths.

Set $q=e^{2\pi i/N}$. Each gate acts by the phase $-q$ on a unit vector $\psi$ and fixes its orthogonal complement, so its matrix in an orthonormal basis is $I-(1+q)\psi\psi^*$. These gates come from three reflections $g_1,g_2,g_3$ on $\C^2$, with $g_1=g_3$. The normalized trace of the first pair is $a=q^{-1}\tr(g_1g_2)=(\tr(AB)-1)/(-q)$, and $b=q^{-1}\tr(g_1^2)$.

At this rational point one can see the length bound directly. Normalize a valuation above five by $v(5)=1$, so $v(3/5)=-1$. The entries of every generator and inverse have valuation at least $-1$. Since $v(x+y)\ge\min\{v(x),v(y)\}$, a $k$-letter word has trace of valuation at least $-k$. After $n\ge1$ iterations of $LR$, the trace of the product of the first and third targets has valuation $-2F_{2n}$, where $F_0=0,F_1=1$ are the initial Fibonacci numbers. Projective rescaling of either target by a representing word leaves this valuation unchanged because the determinants are units. The two words therefore require at least $2F_{2n}$ letters in total.

Kolev computed the free-group substitution for this braid \cite[Section~5.1]{Kolev}. Canonical-height theory \cite{Abboud}, combined with the comparison in Appendix~\ref{app:height}, also gives the stretch factor as the pair-cost growth exponent for nonperiodic algebraic trace points under a pseudo-Anosov braid. For our cyclotomic initial trace triples we compute the valuation of each pair trace at each time, obtain lower bounds that remain valid when additional generators are available, and determine when the displayed free words remain shortest in the projective gate group. For periodic and reducible three-braids the costs grow at most linearly (Lemma~\ref{lem:nonpa}). Every target is itself a single reflection. In the two-qubit encoding of Subsection~\ref{sec:geometry}, with free one-qubit gates, the sum of the two individual CNOT minima stays between four and six. Earlier work constructed gates and decomposed circuits from twisted Burau representations \cite[Sections~7--8]{NegamiGates}. Here the growing cost counts letters in one fixed reflection alphabet.

\paragraph{Monodromy and the gate set.}
Analytic continuation around based loops acts on the local solution space of a linear differential system and gives its monodromy representation. Middle convolution arose in Katz's study of rigid local systems \cite{Katz}. In one variable it transforms representations of the fundamental group of a punctured complex line \cite{DR00,DRMiddle}; Haraoka extended the construction to Knizhnik--Zamolodchikov (KZ)-type equations \cite{HaraokaKZ}.

In our setting, $B_3$ arises by moving the three punctures of a plane around one another. Based loops around the three punctures represent the free generators $x_1,x_2,x_3$ of $F_3$. Moving the punctures acts on this basis by Hurwitz moves. For compatible input, the Katz--Long--Moody (KLM) construction gives a representation of $F_3\rtimes B_3$ and hence representations of both $F_3$ and $B_3$ by restriction \cite{HiroeNegami,NegamiKZ}. Restriction to $F_3$ recovers the multiplicative middle convolution of Dettweiler and Reiter. Their construction \cite[Definition~2.1]{DRPainleve} takes the input representation $x_j\mapsto g_j$ to $x_j\mapsto M_j$ even when the input admits no compatible braid action. Each braid thus specifies three target gates. We measure their shortest word lengths in the fixed initial alphabet. Even the pure braid $L^2$ changes an ordered pair trace in our family, ruling out a compatible extension to $F_3\rtimes P_3$ for the prescribed action (calculation after Proposition~\ref{prop:familytraces}).

For a finite complex reflection group, Brou\'e, Malle and Rouquier construct monodromy representations of its braid group that factor through Hecke algebras \cite{BMR}. Jones used traces on Hecke algebras to extract link invariants from braid representations \cite{JonesHecke}. Aharonov, Jones and Landau implemented a unitary braid representation to approximate the Jones polynomial of a braid closure at roots of unity \cite{AJL}.

Monodromy along loops gives a model for quantum gates \cite{GiorgadzeMonodromy}. KZ connections give braid representations \cite{KohnoKZ}. In the Ising-like and Fibonacci examples, connection matrices express the local braiding operations in a common basis \cite[Sections~3.3--3.4 and~4.1]{GuHaghighatLiu}. Myers, Sati and Schreiber interpret KZ monodromy as gate transport \cite{MyersSatiSchreiber}. The correspondence with KZ-type middle convolution via KLM requires a compatible pure-braid representation \cite{NegamiKZ}.

\noindent\begin{minipage}{\linewidth}
\paragraph{Main results.}
Fix an odd $N\ge7$ and an overlap $0<c<1$. The assumptions on the overlap depend on the question.
\begin{center}
\small
\begin{tabular}{@{}p{.22\linewidth}p{.70\linewidth}@{}}
\toprule
Parameters & Conclusion\\
\midrule
Real $c$, $r\ge3$ & Density in $\PU(r)=U(r)/\{e^{i\theta}I_r:\theta\in\mathbb R\}$ for every qutrit gate set in this family and, in higher rank, whenever $(r-1)\tan(\pi/N)<1$ (Theorems~\ref{thm:integrated} and~\ref{thm:highrank}).\\[3pt]
Rational $c$, same ranks & For any two of the first three evolving targets, a finite-time lower bound on the sum of their minimum lengths in all $r$ generators and inverses. For each pseudo-Anosov braid this sum has exponential growth rate equal to its stretch factor (Theorems~\ref{thm:integrated} and~\ref{thm:highrank}).\\[3pt]
Transcendental $c$, \mbox{$r=3$} & Individual shortest lengths for every iteration of $LR$ (Theorem~\ref{thm:sharpwords}(ii)).\\[3pt]
Rational $c$, \mbox{$r=3$} & The same equalities up to a prescribed horizon if the reduced denominator satisfies \eqref{eq:finitehorizoncondition}. The threshold is uniform in $N$ and the numerator (Theorem~\ref{thm:sharpwords}(iii)).\\[3pt]
Fixed rational $c$, \mbox{$r=3$} & Approximation to error $\epsilon$: a uniform logarithmic upper bound for every target and a matching order of growth in the worst case (Corollary~\ref{cor:logapprox}).\\
\bottomrule
\end{tabular}
\end{center}
\end{minipage}\par\medskip

Theorem~\ref{thm:symmetricorbit} describes the orbit of the trace triple; Sections~\ref{sec:mc}--\ref{sec:arithmetic} transfer its valuations to word-length bounds. Section~\ref{sec:family} constructs the qutrit gates, Section~\ref{sec:highrank} treats arbitrary rank, and Section~\ref{sec:geodesicity} determines the shortest word lengths in rank three. Section~\ref{sec:scope} compares exact synthesis and approximation, and Section~\ref{sec:discussion} discusses related work.

Throughout, $[A]$ denotes the class of an invertible matrix modulo nonzero scalars.

\begin{definition}[Hurwitz actions and mixed words]\label{def:hurwitz}
On triples of invertible matrices, the left Hurwitz moves are
\begin{equation}\label{eq:hurwitzLR}
 \LL(A,B,C)=(ABA^{-1},A,C),\qquad
 \RR(A,B,C)=(A,C,C^{-1}BC),
\end{equation}
corresponding to $\sigma_1$ and $\sigma_2^{-1}$. On an adjacent pair, $(A,B)\mapsto(ABA^{-1},A)$ is the inverse of the Artin move $(A,B)\mapsto(B,B^{-1}AB)$ in \cite[Definition~1.3]{NegamiGates}. Maps compose from right to left. A \emph{mixed word in $\LL,\RR$} uses both moves and contains no inverse letters. For a braid $\beta$ and an initial tuple $\boldsymbol A$, its $n$th Hurwitz iterate is $\boldsymbol A_n=\beta^n\boldsymbol A=(A_{1,n},A_{2,n},A_{3,n})$.
\end{definition}

\begin{definition}[Single-qutrit universality]\label{def:universality}
A tuple of unitary $3\times3$ matrices is \emph{single-qutrit universal up to phase} if the subgroup generated by its projective classes is dense in $\PU(3)$. The projective order of $A$ is the order of $[A]$, or equivalently the least $d\ge1$ for which $A^d$ is scalar, when such a $d$ exists.
\end{definition}

\begin{definition}[Rank-two input data and normalized traces]\label{def:ranktwoinput}
Let $E$ be a subfield of $\C$, let $q\in E^\times\setminus\{1\}$, and choose $u_j\in E^2$ and $\ell_j\in(E^2)^*$ with $\ell_j(u_j)=1$. Write $g_j=\rho(x_j)$ for the images of the free generators of $F_3$ on $V=E^2$. Using the ordered-product convention of \cite[Section~2]{NegamiHermitian}, set
\[
 g_j=I_2+(q-1)u_j\ell_j,\qquad
 \boldsymbol g=(g_1,g_2,g_3),\qquad P_{\rm in}=g_1g_2g_3.
\]
Here $\operatorname{rank}(g_j-I_2)=1$; the cyclotomic gate families below also have finite order. The normalized pair-trace point and symmetric trace parameter are
\begin{equation}\label{eq:inputtraces}
 (x,y,z)=q^{-1}\bigl(\tr(g_1g_2),\tr(g_2g_3),\tr(g_1g_3)\bigr),\qquad
 \kappa_{\rm F}=\frac{(q+1)^2}{q}+\frac{(q+1)\tr P_{\rm in}}{q^2}.
\end{equation}
For a Hurwitz iterate, $(x_n,y_n,z_n)$ denotes its normalized pair-trace point.
\end{definition}

\begin{definition}[Admissible local convolution parameters]\label{def:admissible}
For the input data of Definition~\ref{def:ranktwoinput} and a valuation $v$, a parameter $\lambda\in E^\times$ is \emph{admissible at $v$} if
\begin{equation}\label{eq:admissible}
 v(\lambda)=0,\qquad \lambda\ne1,\qquad \lambda q\ne1,
 \qquad \det(\lambda P_{\rm in}-I_2)\ne0.
\end{equation}
A parameter is \emph{spectrally admissible} if it satisfies the last three conditions, without the requirement $v(\lambda)=0$.
\end{definition}

\section{The \texorpdfstring{$B_3$}{B3}-orbit}\label{sec:fullorbit}
We describe the $B_3$-orbit of the symmetric trace data used in our gate construction and derive valuation formulas for the word-length lower bounds proved later.

\begin{definition}[Integral braid matrices and dilatation notation]\label{def:braidmatrices}
Set
\begin{equation}\label{eq:tropicalmatrices}
 Q_{\LL}=\begin{pmatrix}1&0\\1&1\end{pmatrix},\qquad
 Q_{\RR}=\begin{pmatrix}1&1\\0&1\end{pmatrix},\qquad
 Q_w=Q_{e_1}\cdots Q_{e_k}\quad(w=e_1\cdots e_k),
\end{equation}
using the inverse matrices for inverse letters. Write $Q(\beta)$ for the matrix associated with a braid $\beta$, once we have shown that the assignment is well defined. For a pseudo-Anosov braid $\beta$, $\lambda_\beta>1$ denotes its dilatation (stretch factor), and $h(\beta)=\log\lambda_\beta$ its topological entropy (the infimum over representatives of the mapping class).
\end{definition}

\begin{prop}[The three-braid matrix model]\label{prop:braidmatrices}
The assignments $Q(\sigma_1)=Q_{\LL}$ and $Q(\sigma_2^{-1})=Q_{\RR}$ define a homomorphism $Q:B_3\to SL_2(\mathbb Z)$ whose projectivization identifies $B_3/Z(B_3)$ with $PSL_2(\mathbb Z)$. For every pseudo-Anosov $\beta$, the spectral radius of $Q(\beta)$ equals $\lambda_\beta$. Every mixed word $w$ in $\LL,\RR$ is pseudo-Anosov, $Q_w$ is strictly positive, and its Perron eigenvalue equals $\lambda_w$.
\end{prop}

\begin{proof}
With the assignments $\sigma_1\mapsto Q_{\LL}$ and $\sigma_2\mapsto Q_{\RR}^{-1}$ these are the reduced Burau matrices at $-1$ up to basis and sign. In the standard presentation $B_3=\langle s,t\mid s^2=t^3\rangle$, with $s=\sigma_1\sigma_2\sigma_1$ and $t=\sigma_1\sigma_2$, the central element $s^2$ generates the center. The quotient is $C_2*C_3$, where $C_m$ denotes a cyclic group of order $m$. Direct multiplication gives $Q(s)=S$ and $Q(t)=TS$ for the modular generators $S=\left(\begin{smallmatrix}0&-1\\1&0\end{smallmatrix}\right)$ and $T=\left(\begin{smallmatrix}1&1\\0&1\end{smallmatrix}\right)$. The standard presentation of $PSL_2(\mathbb Z)$ in the generators $[S],[TS]$ gives the asserted isomorphism. The stretch-factor formula for the reduced Burau representation on three strands is standard \cite[Lemma~4.3(a) and Theorem~5.1]{BandBoyland}.

A product containing both positive shears has strictly positive integral entries, determinant one, and trace at least three. It is therefore hyperbolic, and its spectral radius is its Perron eigenvalue.
\end{proof}

\begin{definition}[The initial trace triple and its slope coordinates]\label{def:symmetric-seed}
In the gate family the first and third normalized input matrices coincide, so the normalized matrices $U_j=q^{-1/2}g_j$ give equal $12$ and $23$ traces $a$ and the trace $b=\tr(U_1^2)$ for the $13$ pair. Every pair trace along this specialized orbit is consequently the trace of a word in $U_1,U_2$. If $t$ is the common single trace, then $t^2=b+2$ and Cayley--Hamilton gives the total trace $d=t(a-1)$ (Proposition~\ref{prop:familytraces}). This explains the coefficient $t^2+td=(b+2)a$. For the polynomial orbit, take $a,b$ as independent indeterminates and put
\[
 \kappa_{\rm F}=(b+2)a,\qquad \mathcal E=b^2-(b+2)a(a+b),\qquad s_0=(a,a,b).
\]
The polynomial trace action is
\begin{equation}\label{eq:traceLR}
 \LL(x,y,z)=(x,z,\kappa_{\rm F}-xz-y),\qquad
 \RR(x,y,z)=(z,y,\kappa_{\rm F}-yz-x).
\end{equation}
It preserves $x^2+y^2+z^2+xyz=\kappa_{\rm F}(x+y+z)+\mathcal E$ and defines a $B_3$-action. For $\beta\in B_3$, write
\[
 (x_\beta,y_\beta,z_\beta)=\beta s_0,\qquad
 \binom{r_\beta}{t_\beta}=Q(\beta)\binom{1}{-1}.
\]
The same notation is used after specializing $a,b$ to a field of characteristic zero.
\end{definition}

\begin{maintheorem}[The $B_3$-orbit parametrized by rational slopes]\label{thm:symmetricorbit}
For the indeterminate data of Definition~\ref{def:symmetric-seed}, all three coordinates belong to $\mathbb Z[a,b]$, and for every $\beta\in B_3$,
\begin{equation}\label{eq:globaldegrees}
 \bigl(\deg_a x_\beta,\deg_a y_\beta,\deg_a z_\beta\bigr)
 =\bigl(|r_\beta|,|t_\beta|,|r_\beta+t_\beta|\bigr).
\end{equation}
Every coordinate of positive $a$-degree has leading coefficient $1$ or $-1$ in $\mathbb Z[b]$; a coordinate of degree zero is $b$. The assignment
\[
 \pm(r_\beta,t_\beta)\longmapsto\beta s_0
\]
is a well-defined $B_3$-equivariant bijection from primitive integer vectors modulo simultaneous sign, equivalently $\mathbb P^1(\Q)$, to the polynomial orbit. The stabilizer of $s_0$ is the inverse image of the cyclic subgroup generated by $Q(\RR^{-1}\LL\RR)$ in $PSL_2(\mathbb Z)$.
\end{maintheorem}

\begin{proof}[Proof of Theorem~\ref{thm:symmetricorbit}]
The maps \eqref{eq:traceLR} and their inverses
\[
 \LL^{-1}(x,y,z)=(x,\kappa_{\rm F}-xy-z,y),\qquad
 \RR^{-1}(x,y,z)=(\kappa_{\rm F}-xy-z,y,x)
\]
have integer coefficients in $a,b$. Direct substitution verifies the braid relation for $\LL,\RR^{-1}$ and the identities at the initial point
\begin{equation}\label{eq:seedidentities}
 \begin{gathered}
 \LL s_0=(a,b,a),\quad \RR s_0=(b,a,a),\quad
 \RR\LL s_0=\LL s_0,\quad \LL\RR s_0=\RR s_0,\\
 \LL^2s_0=\RR^2s_0=(a,a,z_*),\qquad z_*=(b+2)a-a^2-b.
 \end{gathered}
\end{equation}
In particular the latter point has $a$-degrees $(1,1,2)$ and leading coefficients $(1,1,-1)$; its slope vector is $(1,1)^T$.

Suppose a point has degree triple $(r,t,r+t)$ with $r,t>0$ and all three leading coefficients in $\{1,-1\}$. In the last coordinate $\kappa_{\rm F}-xz-y$ of its $\LL$-image, the product has degree $2r+t$, strictly larger than $1=\deg_a\kappa_{\rm F}$ and $t=\deg_a y$. Its leading coefficient is therefore again $1$ or $-1$. The same argument applies to $\RR$, giving the degree rules
\[
 (r,t,r+t)\overset{\LL}{\longmapsto}(r,r+t,2r+t),\qquad
 (r,t,r+t)\overset{\RR}{\longmapsto}(r+t,t,r+2t).
\]
The subtractive Euclidean algorithm obtains every coprime positive pair from $(1,1)$ by these two shears.

Put $\mathsf C=\RR^{-1}\LL$. Direct substitution gives
\[
 \mathsf C(x,y,z)=(y,z,x),\qquad Q(\mathsf C)\binom rt=\binom{-t}{r+t}.
\]
Thus both actions cyclically permute the three quantities in \eqref{eq:globaldegrees}. Powers of $Q(\mathsf C)$ move every primitive vector, modulo sign, into the closed nonnegative quadrant. The boundary directions $(1,0)^T$ and $(0,1)^T$, considered modulo simultaneous sign, are represented by $\LL s_0$ and $\RR s_0$, respectively. For these points, \eqref{eq:seedidentities} gives the degree formula and the degree-zero coordinate $b$. Explicitly, a rotated interior direction has representative $\mathsf C^k w\LL^2$ with $w$ a possibly empty word in $\LL,\RR$; a rotated boundary direction has representative $\mathsf C^k\LL$ or $\mathsf C^k\RR$. These representatives give the asserted degrees in every direction.

To see why the choice of representative does not matter, note that the stabilizer of $(1,-1)^T$ modulo sign in $PSL_2(\mathbb Z)$ is generated by $Q(\mathsf S)$, with $\mathsf S=\RR^{-1}\LL\RR$: indeed $Q(\RR)(1,-1)^T=-e_2$, whose projective stabilizer consists of the lower unipotents $Q(\LL)^k$. The identity $\LL\RR s_0=\RR s_0$ shows that $\mathsf S$ fixes $s_0$ exactly. Also $\mathsf C^3=(\sigma_2\sigma_1)^3$ generates the braid center and acts trivially on the coordinates. The full preimage of the slope stabilizer consequently fixes $s_0$. If two braids have the same slope modulo sign, their quotient belongs to that preimage, so their polynomial points coincide.

Finally the triple $(|r|,|t|,|r+t|)$ determines $(r,t)$ modulo simultaneous sign. Distinct primitive directions therefore have distinct polynomial degree triples, hence distinct polynomial points. This proves the bijection and the asserted stabilizer.
\end{proof}

\subsection{Valuations after arithmetic specialization}

We now substitute values from a number field for the indeterminates $a,b$, assuming $v(a)=-m<0$ and $v(b)=0$ at a non-Archimedean place. We compute the valuation of each trace coordinate along the $B_3$-orbit. These valuations will give lower bounds for the pair cost in a fixed gate set.

\begin{corollary}[Exact valuations along the $B_3$-orbit]\label{cor:globalvaluations}
Suppose $a,b$ belong to a number field with a non-Archimedean valuation satisfying $v(a)=-m<0$ and $v(b)=0$. Then, for every $\beta\in B_3$,
\begin{equation}\label{eq:globalweights}
 \bigl(-v(x_\beta),-v(y_\beta),-v(z_\beta)\bigr)
 =m\bigl(|r_\beta|,|t_\beta|,|r_\beta+t_\beta|\bigr).
\end{equation}
The specialized trace orbit remains $B_3$-equivariantly in bijection with $\mathbb P^1(\Q)$, with the same stabilizer as in Theorem~\ref{thm:symmetricorbit}.
\end{corollary}

\begin{proof}
A coordinate of degree $d>0$ has the form $\epsilon a^d+\sum_{j<d}c_j(b)a^j$, with $\epsilon\in\{1,-1\}$ and $c_j\in\mathbb Z[b]$. Since $v(c_j(b))\ge0$, every lower term has valuation at least $-jm>-dm$. Hence the leading term has uniquely smallest valuation, giving $-v=md$. A degree-zero coordinate is $b$ and has valuation zero. Finally $(|r|,|t|,|r+t|)$ determines $(r,t)$ up to simultaneous sign, so distinct slope classes remain distinct after specialization. The polynomial bijection therefore specializes injectively, and its stabilizer is unchanged.
\end{proof}

\begin{corollary}[Recovering the stretch factor from two iterates]\label{cor:twoiterates}
Fix an initial trace triple $s_0=(a,a,b)$ and a valuation $v$ as in Corollary~\ref{cor:globalvaluations}. For a pseudo-Anosov three-braid $\beta$, let $z_n$ be the third coordinate of $\beta^n s_0$ and put $d_n=-v(z_n)$. Then $d_1>0$ and
\[
 T=\frac{d_2}{d_1}=|\tr Q(\beta)|,\qquad
 \lambda_\beta=\frac{T+\sqrt{T^2-4}}{2}.
\]
\end{corollary}
\begin{proof}
Put $Q=Q(\beta)$ and $u=(1,-1)^T$. Equation~\eqref{eq:globalweights} gives $d_n=m|\det(u,Q^nu)|$. Since $Q$ is hyperbolic in $SL_2(\mathbb Z)$, its eigenlines have irrational slope, so $\det(u,Qu)\ne0$. By Cayley--Hamilton, $Q^2=(\tr Q)Q-I$, hence $\det(u,Q^2u)=(\tr Q)\det(u,Qu)$. Taking absolute values gives the ratio, and Proposition~\ref{prop:braidmatrices} identifies the spectral radius of $Q$ with $\lambda_\beta$.
\end{proof}

At a place above a prime dividing the reduced denominator of $c$ in the cyclotomic family, $v(a)=-m<0$ and $v(b)=v(b+2)=0$, whence $v(\kappa_{\rm F})=-m$ and $v(\mathcal E)=-2m$. The nonnegative-parameter hypothesis of \cite[Theorem~4.10]{JangTropical} therefore fails. At the initial point several terms of least valuation can cancel. The next example shows why the initial valuation triple alone does not determine its successor.

\begin{example}[The initial valuation triple does not determine its lift]\label{ex:boundarylift}
Over $\Q$ with $v_7(7)=1$, take $a=1/7$, $b=5/2$, so $\kappa_{\rm F}=9/14$ and $\mathcal E=223/49$. Both
\[
 s_0=(1/7,1/7,5/2),\qquad s_{\rm ex}=(-17/7,10/7,5/2)
\]
lie on this same cubic and have valuation triple $(-1,-1,0)$. But direct substitution gives
\[
 \RR s_0=(5/2,1/7,1/7),\qquad \RR s_{\rm ex}=(5/2,10/7,-1/2),
\]
with valuation triples $(0,-1,-1)$ and $(0,-1,0)$, respectively. Thus the algebraic identities at the initial point, rather than its initial valuation triple alone, are needed to obtain the formula on the $B_3$-orbit.
\end{example}

\subsection{Relation to Farey polynomials}
General trace-degree formulas have classical precedents \cite{BrownTrace}, and rational slopes and denominator-degree formulas occur for elliptic Farey polynomials \cite[Section~2.3]{FareyWords}. To compare the latter with our specialization, in the notation of that paper set $\alpha=\beta=\eta$, $\eta^2=q$, $b=q+q^{-1}$, and substitute $z=a-b$ into their standard pair of generators. For a rational slope $s$, let $\Phi_s$ be the trace of its Farey word after this substitution \cite[Section~2.3]{FareyWords}. Put $C_b=b+2$. For adjacent Farey fractions $u=p_1/q_1,v=p_2/q_2$ in $[0,1]$ (so $|p_1q_2-p_2q_1|=1$), let $w,w'$ be the opposite vertices of the two incident Farey triangles. With the two local parameters equal, the constant in each parity case is $4+2(q+q^{-1})=2C_b$. Thus both cases of \cite[Theorem~4.4]{FareyWords} specialize to $\Phi_w+\Phi_{w'}=2C_b-\Phi_u\Phi_v$, with initial values
\[
 \Phi_{0/1}=C_b-a,\qquad\Phi_{1/1}=a,\qquad\Phi_{1/0}=2.
\]
With $z_*$ as in \eqref{eq:seedidentities}, put $u_*=\kappa_{\rm F}-az_*-a$ and $w_*=\kappa_{\rm F}-u_*z_*-a$. These are coordinates reached by applying $\LL$ and then $\RR$ to $(a,a,z_*)$. The Farey recurrence gives the polynomial identities
\begin{equation}\label{eq:fareycomparison}
 \begin{aligned}
 z_*&=C_b-\Phi_{1/2},\qquad u_*=C_b-\Phi_{2/3},\\
 w_*&=\Phi_{3/5}-(b+2)(a-1)(a-b)(a-2).
 \end{aligned}
\end{equation}
The nonzero correction in the third identity prevents extending the first two identities to a direct identification of the whole polynomial family by the same rule. It has lower degree, so this comparison does not rule out a derivation of degree statements through more general substitutions, such as those treated in \cite[Corollary~4.6]{FareyWords}. Theorem~\ref{thm:symmetricorbit} supplies a direct proof for the specialization $\kappa_{\rm F}=(b+2)a$, including its stabilizer and unit leading coefficients; Corollary~\ref{cor:globalvaluations} then gives the required arithmetic conclusion.

\section{Convolution matrices and common trace dynamics}\label{sec:mc}

Middle convolution constructs operators on a direct sum of copies of the input space $V$ and lets them act on the quotient by an invariant subspace $K+L(\lambda)$ (Definition~\ref{def:mcoperators}). Under suitable hypotheses an invariant Hermitian form on the input induces one on the quotient \cite{NegamiHermitian}. We establish positivity for our gate families below. In the compatible pure-braid setting, the KLM/Haraoka correspondence identifies the algebraic transformation with middle convolution for KZ-type equations \cite[Sections~3--4]{NegamiKZ}. Here we work with its explicit free-generator formulas, identify quotient coordinates, and prove the trace identities used in the subsequent word-length estimates. The braid-equivariance statement below concerns the action on tuples; it does not require the fixed tuple to extend to a full braid representation.

\begin{definition}[Convolution operators and quotient]\label{def:mcoperators}
Let $V=E^d$ for a subfield $E\subset\C$. Fix $g_1,\ldots,g_r\in GL(V)$ and $\lambda\in E^\times\setminus\{1\}$. On $V^{\oplus r}$ the convolution operator $G_j$ equals the identity outside block row $j$, where
\begin{equation}\label{eq:blocks}
 (G_j)_{jk}=\begin{cases}
 \lambda(g_k-I_d),&k<j,\\
 \lambda g_j,&k=j,\\
 g_k-I_d,&k>j.
 \end{cases}
\end{equation}
Set $K=\bigoplus_{k=1}^r\ker(g_k-I_d)$ and $L(\lambda)=\bigcap_{j=1}^r\ker(G_j-I)$. On $K$, the operator $G_j$ scales summand $j$ by $\lambda$ and fixes the other summands. Thus $K$ and $L(\lambda)$ are invariant and intersect trivially. The middle convolution $\MC_\lambda(\boldsymbol g)$ is the tuple induced on $Q_\lambda=V^{\oplus r}/(K+L(\lambda))$. These are the free-generator block formulas of \cite[Definitions~1.4--1.5]{NegamiGates}. For three inputs extending to $F_3\rtimes B_3$, these operators are the restrictions of the KLM representation to the free generators. Our convention reverses the order of the inputs, block coordinates and output labels in \cite[Definition~2.1]{DRPainleve}.
\end{definition}

\begin{definition}[Pairing matrix and convolution matrices for reflection inputs]\label{def:mc-matrices}
Suppose $g_j=I_d+(q-1)u_j\ell_j$, where $q\in E^\times\setminus\{1\}$, $u_j\in V$, $\ell_j\in V^*$, and $\ell_j(u_j)=1$. Define the matrix of pairings $B=(b_{ij})_{1\le i,j\le r}$ by $b_{ij}=\ell_i(u_j)$. Put $\Pi(w_1,\ldots,w_r)=(\ell_1(w_1),\ldots,\ell_r(w_r))$. Let $M_j$ equal the identity on $E^r$ outside row $j$, where
\begin{equation}\label{eq:generalM}
 (M_j)_{jk}=\begin{cases}
 \lambda(q-1)b_{jk},&k<j,\\
 \lambda q,&k=j,\\
 (q-1)b_{jk},&k>j.
 \end{cases}
\end{equation}
Write $\boldsymbol M=(M_1,\ldots,M_r)$ and $\qout=\lambda q$.
\end{definition}

\begin{lemma}[Quotient coordinates for reflection inputs]\label{lem:mc}
The matrix $B$ has diagonal entries one and rank at most $d$. The map $\Pi$ identifies $V^{\oplus r}/K$ with $E^r$ and satisfies $\Pi G_j=M_j\Pi$. Moreover,
\[
 L(\lambda)\simeq\ker(\lambda g_1\cdots g_r-I_d).
\]
If this kernel vanishes, the matrices $M_j$ represent $\MC_\lambda(\boldsymbol g)$ on $E^r$. If also $\qout\ne1$, each is a reflection with special eigenvalue $\qout$.
\end{lemma}
\begin{proof}
The matrix $B$ represents the composite $E^r\to V\to E^r$ given by $e_j\mapsto u_j$ and $v\mapsto(\ell_i(v))_i$. Its diagonal entries are $\ell_i(u_i)=1$. Each $\ell_j$ is surjective and $\ker\ell_j=\ker(g_j-I_d)$, so $\Pi$ is surjective with kernel $K$. Applying $\ell_j$ to row $j$ of \eqref{eq:blocks} and using $\ell_j(g_k-I_d)=(q-1)b_{jk}\ell_k$ gives the intertwining identity.

For $w\in L(\lambda)$, subtraction of consecutive fixed-vector equations gives $(\lambda-1)(w_j-g_{j+1}w_{j+1})=0$ for $j<r$. Substituting $w_j=g_{j+1}\cdots g_rw_r$ into the first row leaves $(\lambda g_1\cdots g_r-I_d)w_r=0$. Conversely these substitutions reconstruct a common fixed vector from every vector in the kernel. Thus projection to $w_r$ gives the asserted isomorphism. In particular the general dimension formula is
\[
 \dim Q_\lambda=\sum_{j=1}^r\operatorname{rank}(g_j-I_d)-\dim\ker(\lambda g_1\cdots g_r-I_d)
\]
as in \cite[Theorem~2.4(i)]{DRMiddle}. For the reflection inputs the sum is $r$. When the kernel vanishes, $\Pi$ gives the claimed quotient coordinates. If $\qout\ne1$, the only nonzero row of $M_j-I_r$ has diagonal entry $\qout-1$, giving the reflection assertion \cite[Definition~2.9]{BeukersHeckman}.
\end{proof}

For $r>d$, when $\lambda\ne1$ and $\det(\lambda g_1\cdots g_r-I_d)\ne0$, the quotient has dimension $r$ and the construction increases rank. A positive invariant Hermitian form permits a common orthonormal basis; Section~\ref{sec:highrank} proves positivity and universality for a concrete family. For the trace dynamics we now take three rank-two inputs, so the three pair traces are coordinates for the Hurwitz action.

\begin{corollary}[Trace identities for three reflections]\label{cor:mctraces}
For $d=2$, $r=3$ and a spectrally admissible $\lambda$ as in Definition~\ref{def:admissible}, set $P_{\rm in}=g_1g_2g_3$ and $P_{\rm out}=M_1M_2M_3$. Then
\begin{equation}\label{eq:mctraces}
 \tr(M_iM_j)=1+\lambda\tr(g_ig_j)\quad(i\ne j),\qquad
 \det(\zeta I_3-P_{\rm out})=(\zeta-\lambda)\det(\zeta I_2-\lambda P_{\rm in}).
\end{equation}
\end{corollary}
\begin{proof}
By Lemma~\ref{lem:mc}, $\operatorname{rank}B\le2$ and hence $\det B=0$.
Set $h=q-1$,
$S_2=b_{12}b_{21}+b_{13}b_{31}+b_{23}b_{32}$ and
$C_3=b_{12}b_{23}b_{31}$. Direct multiplication gives
\[
 \tr(g_ig_j)=2q+h^2b_{ij}b_{ji}\quad(i\ne j),\qquad
 p:=\tr P_{\rm in}=3q-1+h^2S_2+h^3C_3.
\]
The off-diagonal products in \eqref{eq:generalM} give the first identity of \eqref{eq:mctraces}. Expanding the characteristic polynomial of $P_{\rm out}$ gives
\[
 \det(\zeta I_3-P_{\rm out})-(\zeta-\lambda)(\zeta^2-\lambda p\zeta+\lambda^2q^3)
 =-\lambda^2\zeta h^3\det B=0,
\]
which proves the second. For the explicit matrices \eqref{eq:generalM}, this calculation uses only $d=2$, $r=3$ and $\operatorname{rank}B\le2$. The spectral exclusions ensure that these matrices describe the quotient reflections.
\end{proof}

\begin{definition}[Determinant normalization and Fricke surface]\label{def:fricke}
Adjoin a square root $\eta^2=q$ when necessary and define
\[
 U_j=\eta^{-1}g_j,\qquad t=(q+1)/\eta,\qquad
 d=\tr P_{\rm in}/\eta^3,\qquad \mathcal E=4-3t^2-d^2-t^3d.
\]
For these parameters, the symmetric Fricke cubic, the relative $\mathrm{SL}_2$-character variety of the four-punctured sphere with these peripheral traces, is the affine surface
\begin{equation}\label{eq:fricke}
 S_{\kappa_{\rm F},\mathcal E}:\quad
 x^2+y^2+z^2+xyz-\kappa_{\rm F}(x+y+z)-\mathcal E=0.
\end{equation}
For a reflection tuple $\boldsymbol A=(A_1,A_2,A_3)\in GL_3(E)^3$ with common nontrivial eigenvalue $\qout\ne1$, its normalized pair traces are
\[
 \tau_{ij}^{(\qout)}(\boldsymbol A)
 =\frac{\tr(A_iA_j)-1}{\qout}\qquad(i\ne j).
\]
\end{definition}

\begin{prop}[Fricke relation and trace action]\label{prop:fricke-action}
The matrices $U_j$ lie in $SL_2$, have common trace $t$, and satisfy $\tr(U_1U_2U_3)=d$ and $\kappa_{\rm F}=t^2+td$.
The normalized input pair traces lie on $S_{\kappa_{\rm F},\mathcal E}$. Both Hurwitz moves preserve $t,d,\kappa_{\rm F},\mathcal E$, and their actions on these pair traces are the polynomial maps \eqref{eq:traceLR}.
\end{prop}
\begin{proof}
The determinant and trace assertions follow directly from the normalization and \eqref{eq:inputtraces}. The Fricke relation is \cite[Section~1.1]{CantatLoray} with the three local traces equal to $t$.
Both Hurwitz moves preserve the total product and permute the individual conjugacy classes, so all four parameters remain fixed. The elementary identities $C^{-1}=tI-C$ and $CBC=yC-B^{-1}$ for the normalized $SL_2$ matrices give \eqref{eq:traceLR}.
\end{proof}

\begin{lemma}[Preservation of normalized pair traces under the braid action]\label{lem:transport}
For a spectrally admissible $\lambda$, the normalized input and output pair traces agree for every word in the Hurwitz moves and their inverses and every pair $i\ne j$:
\begin{equation}\label{eq:transport}
 q^{-1}\tr(g_{i,w}g_{j,w})
 =\tau_{ij}^{(\lambda q)}(\boldsymbol M_w)
 =\frac{\tr(M_{i,w}M_{j,w})-1}{\lambda q}.
\end{equation}
They follow \eqref{eq:traceLR} with the same $\kappa_{\rm F}$.
\end{lemma}
\begin{proof}
A Hurwitz move preserves the ordered total product and permutes the individual conjugacy classes. Thus every transformed input satisfies the same spectral exclusions. Apply Corollary~\ref{cor:mctraces} to that input to obtain the normalization in \eqref{eq:transport}. The block row \eqref{eq:blocks} uses the order of \cite[Definition~2.1]{DRMiddle}: the factor $\lambda$ multiplies the blocks to the left of the diagonal. In the notation of \cite[Theorem~2.4(iv)]{DRMiddle}, $\LL=Q_1$ and $\RR=Q_2^{-1}$ by \eqref{eq:hurwitzLR}. Part~(iv) applies to any tuple of invertible matrices and any nonzero $\lambda$. It gives simultaneous conjugacy for every braid, so all corresponding pair traces agree. Proposition~\ref{prop:fricke-action} now gives the common trace action. For a related equivariant correspondence of reflection and rank-two trace data, see \cite[Section~2, equation~(17) and Theorem~1]{Boalch}.

\end{proof}

\section{Word growth from a forward-invariant region in valuation space}\label{sec:arithmetic}
\begin{definition}[Valuation weights and the invariant region]\label{def:chamber}
For a non-Archimedean valuation $v$ of $E$, extended by $v(0)=+\infty$, the weights of a point $(x,y,z)$ are $X=-v(x)$, $Y=-v(y)$, and $Z=-v(z)$. The region with parameter $\kappa_{\rm F}$ consists of points satisfying
\begin{equation}\label{eq:chamber}
 X>0,\qquad Y>0,\qquad Z=X+Y>-v(\kappa_{\rm F}).
\end{equation}
\end{definition}
\begin{prop}[Exact tropical dynamics]\label{prop:chamber}
The region \eqref{eq:chamber} is forward-invariant under both moves in \eqref{eq:traceLR}. In this region their action on $(X,Y)^T$ is exactly $Q_{\LL}$ and $Q_{\RR}$, respectively. Let $w$ be a word in $L,R$ with no inverse letters. If its orbit enters the region at time $n_0$, then
\begin{equation}\label{eq:weightgrowth}
 \binom{-v(x_n)}{-v(y_n)}
 =Q_w^{n-n_0}\binom{X_{n_0}}{Y_{n_0}},\qquad
 -v(z_n)=-v(x_n)-v(y_n)\quad(n\ge n_0).
\end{equation}
For a mixed word $w$ in $\LL,\RR$, all three weights are asymptotic to positive constants times $\lambda_w^n$.
\end{prop}
\begin{proof}
Put $Z=X+Y$. For $\LL$, the product $xz$ has weight $X+Z$, strictly greater than $Y$ and $-v(\kappa_{\rm F})$; hence it is the unique term of smallest valuation. Thus
\[
 (X,Y,Z)\overset{\LL}{\longmapsto}(X,X+Y,2X+Y).
\]
Similarly $\RR$ sends it to $(X+Y,Y,X+2Y)$. Both preserve the region and give \eqref{eq:weightgrowth}. A mixed product $Q_w$ has positive integer entries, determinant one and trace at least three. Perron--Frobenius proves the asymptotics.
\end{proof}

\begin{prop}[Zariski density from one place]\label{prop:zariski}
If the forward trace orbit of a mixed word $w$ in $\LL,\RR$ enters the region, it is Zariski dense in the complex surface \eqref{eq:fricke}.
\end{prop}
\begin{proof}
The eigenvalues of $Q_w$ are $\lambda_w,\lambda_w^{-1}$. Its integer trace is at least three, so the discriminant $(\tr Q_w)^2-4$ is not a square. A rational Perron eigenvector would have rational eigenvalue; therefore its slope is irrational.
For a nonzero polynomial $F(U,V)=\sum c_{ij}U^iV^j\in E[U,V]$, the valuations of its terms at $(x_n,y_n)$ are
\[
 v(c_{ij})-iX_n-jY_n.
\]
The irrational limiting slope of $(X_n,Y_n)$ makes one of the finitely many linear combinations $iX_n+jY_n$ uniquely largest for all large $n$, with a gap growing like $\lambda_w^n$. That term has uniquely smallest valuation even after the fixed $v(c_{ij})$ are included. Thus $F(x_n,y_n)\ne0$ for all large $n$.
A polynomial relation with complex coefficients would descend to $E$: expand its finitely many coefficients in an $E$-linearly independent basis of their span and evaluate at the $E$-rational orbit points. Hence the projected orbit is Zariski dense in $\C^2$.

The discriminant of \eqref{eq:fricke}, as a monic quadratic in $z$, is
\[
 (xy-\kappa_{\rm F})^2-4(x^2+y^2-\kappa_{\rm F} x-\kappa_{\rm F} y-\mathcal E).
\]
Its leading coefficient as a polynomial in $x$ is $y^2-4$, a nonsquare in $\C[y]$. Thus the cubic is irreducible. The orbit closure has dimension at least two by the dense projection, so it is the whole surface.
\end{proof}

\begin{definition}[Shortest projective word costs]\label{def:wordcost}
Fix $\rho_{\boldsymbol A}:F_m\to GL_d(E)$ with $x_j\mapsto A_j$. The cost $\len_{\boldsymbol A}([V])$ is the minimum reduced length of a word $u\in F_m$ satisfying $[\rho_{\boldsymbol A}(u)]=[V]$. Each initial generator or inverse counts as one letter. For a Hurwitz iterate $\boldsymbol A_n=\beta^n\boldsymbol A$, define, for $i\ne j$,
\[
 L^{ij}_{\boldsymbol A,\beta}(n)=\len_{\boldsymbol A}([A_{i,n}])+
 \len_{\boldsymbol A}([A_{j,n}]),\qquad
 D_{\boldsymbol A,\beta}(n)=\sum_{j=1}^m\len_{\boldsymbol A}([A_{j,n}]).
\]
Any free word for a target gives an upper bound. The next lemma bounds the pair cost from below.
\end{definition}

\begin{lemma}[Projective arithmetic lower bound]\label{lem:denominator}
Let $S$ be any finite subset of $GL_d(E)$, where $E$ has a non-Archimedean valuation $v$. Assume $v(\det A)=0$ for all $A\in S$, and let $-a_*$ be a common lower bound for the entry valuations of all letters in $S^{\pm1}$, with $a_*>0$. For any $U,V\in\langle S\rangle$ with $\tr(UV)\ne0$,
\begin{equation}\label{eq:lower}
 -v(\tr(UV))\le a_*\bigl(\len_S([U])+\len_S([V])\bigr).
\end{equation}
\end{lemma}
\begin{proof}
A word of length $\ell$ has entries and trace of valuation at least $-a_*\ell$. Choose shortest words $W_U=b_U U$ and $W_V=b_V V$ representing the two classes. Comparing a nonzero matrix entry on each side gives $b_U,b_V\in E^\times$. The determinants of all four matrices have valuation zero, so $d\,v(b_U)=d\,v(b_V)=0$. Thus $\tr(W_UW_V)$ has the same valuation as $\tr(UV)$. Apply the word bound to the concatenation.
\end{proof}

\begin{definition}[Reduced lengths of iterated free generators]\label{def:basedwords}
Let $F_3=\langle x_1,x_2,x_3\rangle$ be the free group with its chosen basis, and let $|W|$ denote reduced based word length. For a three-braid $\beta$, write $\alpha_\beta$ for the substitution determined by $(\beta\boldsymbol A)_j=\rho_{\boldsymbol A}(\alpha_\beta(x_j))$, where $\rho_{\boldsymbol A}(x_j)=A_j$. This uses the Hurwitz convention of Definition~\ref{def:hurwitz}. Define
\[
 \mathfrak b_n(\beta)=\max_{1\le j\le3}|\alpha_\beta^n(x_j)|.
\]
\end{definition}

\begin{lemma}[Free-group upper exponent]\label{lem:upper}
For a pseudo-Anosov three-braid $\beta$ of dilatation $\lambda_\beta$,
\begin{equation}\label{eq:uppergrowth}
 \limsup_{n\to\infty}\mathfrak b_n(\beta)^{1/n}\le\lambda_\beta.
\end{equation}
In every matrix representation, $L^{ij}_{\boldsymbol A,\beta}(n)\le2\mathfrak b_n(\beta)$.
\end{lemma}
\begin{proof}
Choose the basepoint on the boundary and representatives fixing that boundary pointwise. Kolev's definition of automorphism growth uses reduced based words, and his equation~(27) identifies it with $\limsup_n\mathfrak b_n(\beta)^{1/n}$; Bowen's inequality then bounds it by $\exp h(f)$ for every representative $f$ \cite[equations~(1), (26)--(27), and footnote~1]{Kolev}. The boundary-fixed Thurston representative has a zero-entropy collar and a pseudo-Anosov part of entropy $\log\lambda_\beta$ \cite[Theorem~2.1 and Section~4]{BandBoyland}. Thus taking the infimum over representatives proves \eqref{eq:uppergrowth}. Evaluating the reduced free words proves the pair bound. This gives $O_\epsilon((\lambda_\beta+\epsilon)^n)$ for every $\epsilon>0$; the explicit finite-time bound for $\LL\RR$ is proved separately in Theorem~\ref{thm:fibonacci}.
\end{proof}

\begin{lemma}[At most linear growth for non-pseudo-Anosov three-braids]\label{lem:nonpa}
For every periodic or non-periodic reducible $\beta\in B_3$, there is a constant $C_\beta>0$ such that
\[
 \mathfrak b_n(\beta)\le C_\beta(n+1)\qquad(n\ge0).
\]
Consequently, the individual exact word costs, their pair sums, and their total are $O_\beta(n+1)$ in every fixed matrix alphabet. These upper bounds remain valid when further generators are allowed.
\end{lemma}
\begin{proof}
Write $z=\Delta^2=(\sigma_1\sigma_2)^3$. It acts by simultaneous conjugation by $P=ABC$ in our Hurwitz convention. On a pair, put $S=AB$; direct iteration gives, for every integer $m$,
\[
 \sigma_1^{2m}(A,B,C)=(S^mAS^{-m},S^mBS^{-m},C).
\]
Thus the images under $\sigma_1^k$ have based word length $O(|k|+1)$. A non-periodic reducible three-braid is conjugate to $\sigma_1^k z^\ell$ for some integers $k\ne0,\ell$ \cite[Introduction and Proposition~3.1]{AguileraThreeBraids}. Since $\sigma_1$ fixes $P$ and commutes with $z$, the $n$th power of this normal form has based word length $O(n+1)$ on each generator. Conjugating by fixed free-group automorphisms changes this bound by only a constant factor.

If $\beta$ is periodic, then $\beta^d=z^\ell$ for some $d>0$ \cite[Section~6]{GonzalezMeneses}. Writing $n=ad+s$ with $0\le s<d$ reduces its action to conjugation by $P^{a\ell}$ followed by one of finitely many fixed substitutions, again giving $O(n+1)$. Evaluating these representing words in the chosen matrices proves all the cost bounds; allowing more letters cannot increase a minimum length.
\end{proof}

\paragraph{The local transfer criterion.}
We use the rank-two data and local admissibility of Definitions~\ref{def:ranktwoinput} and~\ref{def:admissible}, the region of Definition~\ref{def:chamber}, and the fixed-alphabet costs of Definition~\ref{def:wordcost}. Each pair cost is the sum of two individual shortest lengths.

\begin{prop}[Finite-time arithmetic transfer through middle convolution]\label{thm:main}
Take the input data of Definition~\ref{def:ranktwoinput} and a non-Archimedean valuation $v$ with $v(q)=0$. Let $w$ be a mixed word in $\LL,\RR$ whose trace orbit enters the region at a $w$-iterate $n_0\ge0$. For a parameter $\lambda$ admissible at $v$, let $\boldsymbol M=\MC_\lambda(\boldsymbol g)$ in the explicit rank-three realization of Lemma~\ref{lem:mc}. For either initial alphabet $\boldsymbol A\in\{\boldsymbol g,\boldsymbol M\}$ of dimension $d_{\boldsymbol A}$, define its entry-valuation constant by
\begin{equation}\label{eq:alphabetvaluation}
 a_{\boldsymbol A}=\max\bigl(\{0\}\cup
 \{-v((A_j^{\epsilon})_{kl}):1\le j\le3,\ \epsilon\in\{1,-1\},\ 1\le k,l\le d_{\boldsymbol A}\}\bigr).
\end{equation}
Then $a_{\boldsymbol A}>0$. For every $n\ge n_0$, the common trace weights satisfy
\[
 \binom{X_n}{Y_n}=Q_w^{n-n_0}\binom{X_{n_0}}{Y_{n_0}},\qquad Z_n=X_n+Y_n,
\]
and both alphabets satisfy the finite-time bounds
\begin{equation}\label{eq:transferfinite}
 L^{12}_{\boldsymbol A,w}(n)\ge\frac{X_n}{a_{\boldsymbol A}},\qquad
 L^{23}_{\boldsymbol A,w}(n)\ge\frac{Y_n}{a_{\boldsymbol A}},\qquad
 L^{13}_{\boldsymbol A,w}(n)\ge\frac{X_n+Y_n}{a_{\boldsymbol A}}.
\end{equation}
Consequently, for every pair $i\ne j$,
\begin{equation}\label{eq:mainlimit}
 \lim_{n\to\infty}L^{ij}_{\boldsymbol g,w}(n)^{1/n}
 =\lim_{n\to\infty}L^{ij}_{\boldsymbol M,w}(n)^{1/n}=\lambda_w.
\end{equation}
The same limits hold for the total costs $D$. The common normalized trace orbit is Zariski dense in its symmetric Fricke cubic.
\end{prop}

\begin{proof}[Proof of Proposition~\ref{thm:main}]
Lemma~\ref{lem:transport} identifies the two trace dynamics. In the region, the input pair traces are $qx_n,qy_n,qz_n$ and the output pair traces are $1+\lambda qx_n,1+\lambda qy_n,1+\lambda qz_n$. Since $v(q)=v(\lambda)=0$ and all three coordinate valuations are negative, the corresponding negative trace valuations in both ranks are exactly $X_n,Y_n,X_n+Y_n$. Proposition~\ref{prop:chamber} gives their stated matrix recurrence for every $n\ge n_0$.

The constants in \eqref{eq:alphabetvaluation} are finite. If $a_{\boldsymbol A}=0$, every initial letter and inverse is integral at $v$, and hence so is every Hurwitz word and its pair trace. This contradicts the negative pair-trace valuations at $n_0$, so $a_{\boldsymbol A}>0$. Input and output determinants are $q$ and $\lambda q$, respectively, and remain units. Apply Lemma~\ref{lem:denominator} with $a_*=a_{\boldsymbol A}$ to the three exact trace valuations; this proves \eqref{eq:transferfinite} for all $n\ge n_0$.

Perron--Frobenius gives positive multiples of $\lambda_w^n$ for the three weights. The finite-time inequalities and Lemma~\ref{lem:upper} therefore give the pair-cost limits. Bounding a total cost below by any pair and above by $3\mathfrak b_n(w)$ gives its limit. Proposition~\ref{prop:zariski} proves Zariski density.
\end{proof}
Every inequality for a pair sum also bounds the larger individual cost: for example,
\[
 \max\{\len_{\boldsymbol A}([A_{1,n}]),\len_{\boldsymbol A}([A_{3,n}])\}
 \ge\frac{X_n+Y_n}{2a_{\boldsymbol A}}.
\]
The pair trace controls the sum without identifying which individual target carries that cost. Subsection~\ref{sec:fibonacci} gives the explicit Fibonacci estimates for a particular word.

\begin{corollary}[One fixed trace triple detects every pseudo-Anosov braid]\label{cor:allbraids}
Under Corollary~\ref{cor:globalvaluations}, every pseudo-Anosov $\beta\in B_3$ has a Zariski-dense cyclic trace orbit through $s_0$, and all three negative trace-coordinate valuations have $n$th-root limit $\lambda_\beta$ along $\beta^n$.
If the input in Proposition~\ref{thm:main} has normalized trace point $s_0$ and Fricke coefficient $\kappa_{\rm F}=(b+2)a$, then, for every admissible $\lambda$ satisfying \eqref{eq:admissible}, both ranks have
\[
 \lim_{n\to\infty}L^{ij}_{\boldsymbol g,\beta}(n)^{1/n}
 =\lim_{n\to\infty}L^{ij}_{\boldsymbol M,\beta}(n)^{1/n}
 =\lambda_\beta\qquad(i\ne j).
\]
\end{corollary}
\begin{proof}
The hyperbolic integral matrix $Q(\beta)$ has irrational eigenlines. Thus $(1,-1)^T$ is not in its contracting eigenline. None of the linear forms $r,t,r+t$ vanishes on the expanding eigenline. It follows that the three absolute values in \eqref{eq:globalweights}, applied to $\beta^n$, grow as positive constants times $\lambda_\beta^n$. The limiting ratio $|r_{\beta^n}|/|t_{\beta^n}|$ is irrational. The monomial-valuation argument of Proposition~\ref{prop:zariski} proves Zariski density. Finally trace transport, Lemma~\ref{lem:denominator}, and the general pseudo-Anosov upper bound of Lemma~\ref{lem:upper} give the two word-growth limits.
\end{proof}

\section{Universal qutrit gates and their exact synthesis costs}\label{sec:family}
The construction and denominator estimates lead to the Fibonacci bounds and the proof of Theorem~\ref{thm:integrated} in Subsection~\ref{sec:fibonacci}.
\subsection{Construction and universality}
\begin{definition}[Cyclotomic input and output matrices]\label{def:cyclotomic}
For an odd integer $N\ge7$ and a real parameter $0<c<1$, define $q=e^{2\pi i/N}$, $\eta_c=\sqrt{1-c^2}$, and
\begin{equation}\label{eq:familyT}
 v_1=v_3=(1,0)^T,\qquad v_2=(c,\eta_c)^T,\qquad
 g_j=I_2+(q-1)v_jv_j^*.
\end{equation}
Define the associated rank-three matrices by
\begin{equation}\label{eq:familyM}
\begin{aligned}
 M_1&=\begin{pmatrix}-q&c(q-1)&q-1\\0&1&0\\0&0&1\end{pmatrix},&
 M_2&=\begin{pmatrix}1&0&0\\-c(q-1)&-q&c(q-1)\\0&0&1\end{pmatrix},\\
 M_3&=\begin{pmatrix}1&0&0\\0&1&0\\1-q&-c(q-1)&-q\end{pmatrix}.&&
\end{aligned}
\end{equation}
\end{definition}

\begin{definition}[Hermitian form for the cyclotomic family]\label{def:familyform}
For the data of Definition~\ref{def:cyclotomic}, the input form is $H=I_2$. Normalize the quotient form to have unit diagonal by setting
\[
 \alpha=\frac{1-q}{1+q},\qquad
 H_{\rm q}=\begin{pmatrix}1&c\alpha&\alpha\\c\bar\alpha&1&c\alpha\\
 \bar\alpha&c\bar\alpha&1\end{pmatrix}.
\]
\end{definition}

The following theorem summarizes the results for this family. Universality means density of the projective gate group in $\PU(3)$, as in Definition~\ref{def:universality}. We complete the proof of the theorem at the end of Subsection~\ref{sec:fibonacci} after the finite-time estimates have been established.

\begin{maintheorem}[Universality and dynamical synthesis costs]\label{thm:integrated}
Fix an odd integer $N\ge7$, fix a real $0<c<1$, and set $q=e^{2\pi i/N}$. Let $\boldsymbol g$ and $\boldsymbol M$ be the explicit tuples in \eqref{eq:familyT} and \eqref{eq:familyM}, and specialize the initial trace triple by
\begin{equation}\label{eq:familyparams}
 b=q+q^{-1},\qquad a=2+c^2(b-2),\qquad s_0=(a,a,b).
\end{equation}
Part (i) holds for these real parameters. In parts (ii) and (iii) assume additionally that $c=A_c/B_c$ is rational in lowest terms, with $A_c,B_c>0$.
\begin{enumerate}[label=\textup{(\roman*)},leftmargin=*]
\item \textbf{Unitary realization.} The tuple $\boldsymbol M$ is $\MC_{-1}(\boldsymbol g)$. It preserves the explicit positive Hermitian form $H_{\rm q}$ of Definition~\ref{def:familyform}. After one simultaneous unitarization, its three gates have ordinary and projective order $2N$ and generate a dense subgroup of $\PU(3)$. The input projective closure is $\PU(2)$.
\item \textbf{Uniform finite-time bounds.} For any prime $p\mid B_c$ and any extension $v$ to $\Q(q,\sqrt{1-c^2})$ of the valuation $v(p)=1$, put $\delta=\operatorname{ord}_p(B_c)-v(q-1)>0$. Corollary~\ref{cor:globalvaluations} applies with $m=2\delta$. For every braid $\beta$, the output pair costs satisfy
\[
 L^{12}_{\boldsymbol M,\beta}(1)\ge2|r_\beta|,\quad
 L^{23}_{\boldsymbol M,\beta}(1)\ge2|t_\beta|,\quad
 L^{13}_{\boldsymbol M,\beta}(1)\ge2|r_\beta+t_\beta|.
\]
These constants are independent of $N,c$.
\item \textbf{Dynamical word costs.} For every pseudo-Anosov three-braid $\beta$ and every $i\ne j$,
\[
 \lim_{n\to\infty}L^{ij}_{\boldsymbol g,\beta}(n)^{1/n}
 =\lim_{n\to\infty}L^{ij}_{\boldsymbol M,\beta}(n)^{1/n}
 =\lambda_\beta.
\]
The same limits hold for the total costs $D$, and each cyclic trace orbit is Zariski dense in its Fricke surface. For $\phi=\LL\RR=\sigma_1\sigma_2^{-1}$ and $n\ge1$,
\[
 2F_{2n}\le L^{13}_{\boldsymbol M,\phi}(n)\le2F_{2n+2},
\]
where $F_0=0,F_1=1$ and $F_{j+2}=F_{j+1}+F_j$.
\end{enumerate}
\end{maintheorem}

\begin{prop}[Positive form and quotient realization]\label{prop:unitary}
The matrices $g_j$ in Definition~\ref{def:cyclotomic} are order-$N$ unitary reflections over $\Q(q,c,\eta_c)$. The form of Definition~\ref{def:familyform} satisfies $H_{\rm q}>0$ and $M_j^*H_{\rm q}M_j=H_{\rm q}$. The tuple \eqref{eq:familyM} is $\MC_{-1}$ of \eqref{eq:familyT}, and each $M_j$ has ordinary and projective order $2N$.
\end{prop}
\begin{proof}
The rank-one orthogonal projections $v_jv_j^*$ give the eigenvalues $1,q$ of $g_j$, proving the input assertion. Since $\alpha=-i\tan(\pi/N)$, the eigenvalues of $H_{\rm q}$ are $1$ and
$1\pm\sqrt{1+2c^2}\tan(\pi/N)$. Since $c<1$ and $N>6$,
$\sqrt{1+2c^2}\tan(\pi/N)<\sqrt3\tan(\pi/6)=1$.
The operator $\mathsf p_j=e_j(e_j^*H_{\rm q})$ is the $H_{\rm q}$-orthogonal projection onto $\C e_j$, and $M_j=I-(q+1)\mathsf p_j$. This proves unitarity; the eigenvalues of $M_j$ are $1,1,-q$, and $-q$ has order $2N$ because $N$ is odd.
The intertwining identity $\Pi G_j=M_j\Pi$ proved in Lemma~\ref{lem:mc} does not require spectral nonresonance. To verify $L(-1)=0$ without a separate spectral calculation of the input product, let $w$ be fixed by all convolution operators. Then $\Pi w$ is fixed by all $M_j$, so $H_{\rm q}\Pi w=0$ and $w\in\ker\Pi=K$. On $K$ the $j$th convolution operator negates summand $j$ and fixes the others. Thus $w=0$ and $L(-1)=0$. The fixed-space calculation in Lemma~\ref{lem:mc} also shows $\det(P_{\rm in}+I_2)\ne0$.
\end{proof}

\begin{prop}[Projective density]\label{prop:density}
For Definition~\ref{def:cyclotomic}, the closures of the input and output groups, in orthonormal bases for their respective forms, are projectively $\PU(2)$ and $\PU(3)$.
\end{prop}
\begin{proof}
In $\PU(2)\simeq SO(3)$, two rotations of order greater than five with distinct axes generate a dense subgroup. By the classification of finite rotation groups \cite[Theorem~6.12.1]{ArtinRotations}, the polyhedral groups have maximal rotation order five, and in the cyclic and dihedral cases all higher-order rotations share the principal axis. An infinite proper closed subgroup has a one-dimensional identity component: $\mathfrak{so}(3)$, with bracket given by the cross product, has no two-dimensional subalgebra. The subgroup normalizes that circle of rotations, so every rotation in it of order greater than two has the same axis. Unit vectors spanning the input lines have inner product of modulus $c\in(0,1)$, so they represent neither identical nor antipodal points on the Bloch sphere. The rotations have distinct axes and order $N>5$; this proves input density.

For the output, each pair preserves $E_{ij}=\operatorname{span}(e_i,e_j)$ and fixes $E_{ij}^{\perp_{H_{\rm q}}}$ pointwise. The inequalities $0<|(H_{\rm q})_{ij}|<1$ give distinct axes, of projective order $2N>5$. Let $G_{ij}$ be the compact closure of the restricted pair. Its projective image is all of $\PU(E_{ij})$, while its determinant image is finite, generated by $-q$. Hence its Lie algebra lies in $\mathfrak{su}(E_{ij})$ and maps onto the Lie algebra of $\PU(E_{ij})$. These Lie algebras have the same dimension, so $G_{ij}^{\circ}=\SU(E_{ij})$. The full closure therefore contains these plane groups, acting as the identity on $E_{ij}^{\perp_{H_{\rm q}}}$.
Choose an $H_{\rm q}$-orthonormal basis $u_1,u_2,u_3$ such that
\[
 E_{12}=\langle u_1,u_2\rangle,\qquad
 E_{23}=\langle u_1,\gamma u_2+\omega u_3\rangle,
 \quad\gamma\ge0,\quad\omega>0.
\]
The common line is spanned by $u_1$ and phases make the two coefficients real. With matrix units $E_{ij}^{\rm mat}$, put
$A_{ij}=E_{ij}^{\rm mat}-E_{ji}^{\rm mat}$ and
$B_{ij}=i(E_{ij}^{\rm mat}+E_{ji}^{\rm mat})$.
The two Lie algebras contain $A_{12},B_{12}$ and
$\gamma A_{12}+\omega A_{13},\gamma B_{12}+\omega B_{13}$.
They therefore generate $A_{13},B_{13}$; brackets yield all remaining off-diagonal and traceless diagonal directions. Their generated Lie algebra is $\mathfrak{su}(3)$.
\end{proof}

\paragraph{Common unitarization.}
If $S$ is one common unitarizing change of basis, then
\[
 \len_{S\boldsymbol A S^{-1}}([SVS^{-1}])=\len_{\boldsymbol A}([V]).
\]
Indeed, conjugating each letter gives a length-preserving bijection between representing words and preserves equality up to a scalar. A common unitarization therefore does not change any of the exact costs.

\begin{prop}[Trace parameters of the cyclotomic family]\label{prop:familytraces}
The normalized input pair-trace point is $s_0$ of \eqref{eq:familyparams}. Its Fricke coefficients satisfy
\[
 \kappa_{\rm F}=(b+2)a,\qquad
 \mathcal E=2a^2+b^2+a^2b-\kappa_{\rm F}(2a+b).
\]
\end{prop}
\begin{proof}
Direct multiplication gives the normalized pair traces $(a,a,b)$. Choose $\eta=q^{(N+1)/2}$, so $\eta^2=q$. Then $t=\eta+\eta^{-1}$ and $t^2=b+2$. Since $U_1=U_3$, Cayley--Hamilton gives $d=t(a-1)$, and hence $\kappa_{\rm F}=t^2+td=(b+2)a$. Substitution of $s_0$ into the Fricke equation gives the stated value of $\mathcal E$.
\end{proof}

\paragraph{The free-group image and braid transport.}
The distinction between a free-group tuple and a compatible braid representation is visible even for pure braids in this family. An extension to $F_3\rtimes P_3$, where $P_3$ is the pure braid group with the prescribed Artin action, would realize each pure Hurwitz move by simultaneous conjugation and hence preserve every ordered pair trace. However, \eqref{eq:traceLR} and \eqref{eq:familyparams} give
\[
 \LL^2(a,a,b)=\bigl(a,a,b+c^2(1-c^2)(2-b)^2\bigr)\ne(a,a,b).
\]
Here $\LL^2=\sigma_1^2$ is pure and $0<c<1$, $b<2$. Thus the input has no such compatible extension; Lemma~\ref{lem:transport} gives the same obstruction for the output.

\begin{definition}[Denominator-place data]\label{def:denominator-data}
For the denominator estimates, take $c=A_c/B_c\in(0,1)$ rational in lowest terms, with positive numerator and denominator. Let $\Phi_N$ denote the $N$th cyclotomic polynomial and $\varphi$ Euler's totient function. For a prime $p\mid B_c$, let $v$ be an extension to $\Q(q,\eta_c)$ of the valuation normalized by $v(p)=1$. Define
\begin{equation}\label{eq:delta}
 \nu_p=\operatorname{ord}_p B_c,\qquad
 \varepsilon=v(q-1),\qquad \delta=\nu_p-\varepsilon.
\end{equation}
\end{definition}

\begin{lemma}[Denominator-prime arithmetic]\label{lem:familyval}
For every place in Definition~\ref{def:denominator-data},
\[
 \varepsilon=
 \begin{cases}1/\varphi(N),&N\text{ is a power of }p,\\0,&\text{otherwise},\end{cases}
 \qquad \delta>0,
\]
and
\[
 v(a)=v(\kappa_{\rm F})=-2\delta,\qquad v(b)=0.
\]
Every output letter has entry valuations at least $-\delta$; every input letter has entry valuations at least $-(2\nu_p-\varepsilon)$.
\end{lemma}
\begin{proof}
For any nontrivial odd-order root $u$,
$(1+u)(u+u^3+\cdots+u^{N-2})=-1$ when $u^N=1$ and $N$ is odd.
Thus $1+q$ and $1+q^2$ are algebraic units. Consequently
$b=q^{-1}(1+q^2)$ and $b+2=q^{-1}(q+1)^2$ are units, while
$b-2=q^{-1}(q-1)^2$.
The usual cyclotomic valuation gives the stated $\varepsilon$: a root of unity congruent to one has $p$-power order, and for primitive $p^k$-th roots all $q^a-1$, $(a,p)=1$, have the same valuation; their product is $\Phi_{p^k}(1)=p$.
Since $\nu_p\ge1$ and $\varphi(N)\ge6$, $\delta>0$.
The term $c^2(b-2)$ has valuation $-2\delta<0$, so it dominates the constant in $a$. This proves the initial valuations, including at $p=2$.

For the output, $v(c(q-1))=-\delta$ and $v(q-1)=\varepsilon\ge0$. The rank-one inverse formula gives the same bound for the inverse letters. For the input, $\eta_c^2=1-c^2$ implies $v(\eta_c)=-\nu_p$. The entries of $(q-1)v_jv_j^*$ and their inverse counterparts have valuations at least $\varepsilon-2\nu_p$.
\end{proof}

\begin{lemma}[Uniform entry into the invariant region]\label{lem:entry}
At each place specified in Definition~\ref{def:denominator-data}, for any mixed word $w$ in $\LL,\RR$ the orbit of \eqref{eq:familyparams} enters the region by time $2$. There is a positive integral vector $u_w$, depending only on the written word $w$, such that for every place in Lemma~\ref{lem:familyval} and every $n\ge2$,
\begin{equation}\label{eq:uniformweights}
 \binom{-v(x_n)}{-v(y_n)}=2\delta Q_w^{n-2}u_w,\qquad
 -v(z_n)=2\delta(1,1)Q_w^{n-2}u_w.
\end{equation}
\end{lemma}
\begin{proof}
Use the exact identities \eqref{eq:seedidentities}. Writing $m=2\delta$, the point $\LL^2s_0=(a,a,z_*)$ has weights $(m,m,2m)$ by Corollary~\ref{cor:globalvaluations}, so it is in the region. In the infinite applied-letter sequence $ww\cdots$, the first applied symbol occurs again by the second copy of $w$. Before that occurrence, the other symbol fixes the first image; at that occurrence the region is reached. All subsequent letters act by the integral matrices in \eqref{eq:tropicalmatrices}. The weight vector at the end of the second copy is therefore $m u_w$, independently of $N,c,p$. Proposition~\ref{prop:chamber} finishes the proof.
\end{proof}

Lemma~\ref{lem:entry} gives a uniform entry time for mixed words in $\LL,\RR$. The slope formula yields bounds for every braid, including inverse words.
\begin{prop}[Uniform costs on the $B_3$-orbit]\label{prop:uniformallbraid}
For rational $0<c<1$ and every $\beta\in B_3$, the cyclotomic output satisfies
\begin{equation}\label{eq:uniformlower}
 L^{12}_{\boldsymbol M,\beta}(1)\ge2|r_\beta|,\qquad
 L^{23}_{\boldsymbol M,\beta}(1)\ge2|t_\beta|,\qquad
 L^{13}_{\boldsymbol M,\beta}(1)\ge2|r_\beta+t_\beta|.
\end{equation}
These bounds are independent of $N,c$. They also hold for the explicit general-parameter output \eqref{eq:generalM} whenever \eqref{eq:admissible} holds at a place above $p\mid B_c$.
\end{prop}
\begin{proof}
Let $u$ denote the corresponding nonnegative integer $|r_\beta|$, $|t_\beta|$, or $|r_\beta+t_\beta|$. If $u>0$, Corollary~\ref{cor:globalvaluations} gives the normalized input pair trace $\tau_{\rm pair}$ valuation $-2\delta u<0$. Lemma~\ref{lem:transport} gives output trace $1+\lambda q\tau_{\rm pair}$, whose valuation is again $-2\delta u$, since $v(\lambda q)=0$ and the two summands have distinct valuations. Every output letter and inverse has entries of valuation at least $-\delta$: this is Lemma~\ref{lem:familyval} for $\lambda=-1$, and follows from \eqref{eq:generalM} and the rank-one inverse formula for a general admissible local-unit parameter. Lemma~\ref{lem:denominator} therefore gives $\delta L^{ij}\ge2\delta u$. The same positive factor $\delta$ cancels, which makes the resulting bound independent of $N,c$. If $u=0$, the asserted bound is trivial.
\end{proof}
The same argument on the input gives coefficient $2\delta/(2\nu_p-\varepsilon)$ in front of the corresponding slope weight. Middle convolution preserves the normalized pair traces, while reducing the available per-letter denominator bound from $2\nu_p-\varepsilon$ to $\delta$. This is why the output coefficient becomes the uniform constant $2$; its unitary realization also changes from rank two to a universal qutrit gate set.

\subsection{Fibonacci bounds for exact synthesis}\label{sec:fibonacci}
\begin{definition}[The trace map for $\LL\RR$]\label{def:fibonacci-map}
Define $\phi=\LL\RR$, let $f$ denote its induced map on trace coordinates, and define $\Lambda=(3+\sqrt5)/2$. For the initial point $s_0$, define $(x_n,y_n,z_n)=f^n(s_0)$.
\end{definition}

\begin{prop}[Explicit Fibonacci action]\label{prop:fibonaccimap}
The tuple and trace maps of Definition~\ref{def:fibonacci-map} are
\[
 \phi(A,B,C)=(ACA^{-1},A,C^{-1}BC),
\]
\begin{equation}\label{eq:f}
 f(x,y,z)=(z,\kappa_{\rm F}-yz-x,\kappa_{\rm F}-z(\kappa_{\rm F}-yz-x)-y).
\end{equation}
The dilatation of $\phi$ is $\Lambda$.
\end{prop}
\begin{proof}
Compose the two actions in \eqref{eq:hurwitzLR} and \eqref{eq:traceLR}. The matrix $Q_\phi=Q_{\LL}Q_{\RR}$ has characteristic polynomial $\zeta^2-3\zeta+1$, whose larger root is $\Lambda$.
\end{proof}
\begin{theorem}[Uniform finite-time formulas]\label{thm:fibonacci}
In the cyclotomic family, at every place in Lemma~\ref{lem:familyval} and for every $n\ge1$,
\begin{equation}\label{eq:fibov}
 (v(x_n),v(y_n),v(z_n))
 =-2\delta(F_{2n-2},F_{2n-1},F_{2n}).
\end{equation}
The shortest pair costs in the initial alphabets satisfy
\begin{align}
 2F_{2n}&\le L^{13}_{\boldsymbol M,\phi}(n)\le2F_{2n+2},\label{eq:outputfinite}\\
 \frac{2(\nu_p-\varepsilon)}{2\nu_p-\varepsilon}F_{2n}
 &\le L^{13}_{\boldsymbol g,\phi}(n)\le2F_{2n+2}.\label{eq:inputfinite}
\end{align}
In particular the constants in the output bounds are independent of $N,c,p$. Theorem~\ref{thm:sharpwords}(iii) gives equality with the upper bound up to any prescribed finite horizon under an explicit denominator condition.
\end{theorem}
\begin{proof}
The identity $f(s_0)=(b,a,a)$ establishes \eqref{eq:fibov} for $n=1$. If it holds at $n$, the product $y_nz_n$ has valuation $-2\delta F_{2n+1}$, strictly smaller than those of $\kappa_{\rm F},x_n$. Hence it determines $v(y_{n+1})$. The same argument with $z_ny_{n+1}$ gives $v(z_{n+1})=-2\delta F_{2n+2}$, and $x_{n+1}=z_n$.

The transported pair traces are $qz_n$ and $1-qz_n$ in ranks two and three. Their common valuation is $-2\delta F_{2n}$. Lemma~\ref{lem:denominator}, with the two letter bounds in Lemma~\ref{lem:familyval}, proves the lower estimates.
For the upper bounds, unreduced Hurwitz substitutions give a length vector satisfying
\[
 \boldsymbol u_{n+1}=\begin{pmatrix}2&0&1\\1&0&0\\0&1&2\end{pmatrix}\boldsymbol u_n,
 \qquad\boldsymbol u_0=(1,1,1)^T.
\]
This free-group example also appears in \cite[Section~5.1]{Kolev}. For $n\ge1$, induction gives
\[
 \boldsymbol u_n=(2F_{2n+1}-1,\ 2F_{2n-1}-1,\ 2F_{2n}+1)^T.
\]
The first and third entries sum to $2F_{2n+2}$, proving the bounds in either representation.
\end{proof}

\begin{proof}[Completion of the proof of Theorem~\ref{thm:integrated}]
Part (i) follows from Propositions~\ref{prop:unitary} and~\ref{prop:density}. The denominator calculation in Lemma~\ref{lem:familyval} gives $v(a)=-2\delta$ and $v(b)=0$, and Proposition~\ref{prop:familytraces} gives the required Fricke coefficient; Corollary~\ref{cor:globalvaluations} and Proposition~\ref{prop:uniformallbraid} prove (ii). Corollary~\ref{cor:allbraids} proves the pair limits in (iii). A total cost is bounded below by any pair cost and above by $3\mathfrak b_n(\beta)$, giving its limit. Theorem~\ref{thm:fibonacci} supplies the finite-time Fibonacci bounds.
\end{proof}

\subsection{Arithmetic of the qutrit gate set}\label{sec:qarith}
At a place above $p\mid B_c$, the determinant $1-\lambda\tr P_{\rm in}+\lambda^2q^3$ is nonzero whenever $v(\lambda)=0$: by \eqref{eq:inputtraces} and Lemma~\ref{lem:familyval}, its middle term has valuation $-2\delta<0$, and its other terms are units. This argument also applies after extending the valuation to include $\lambda$. The remaining spectral conditions in \eqref{eq:admissible} are $\lambda\ne1$ and $\lambda q\ne1$.

At the chosen embedding, where $q=e^{2\pi i/N}$ and $\eta_c=\sqrt{1-c^2}>0$, the determinant-normalized input matrices lie in $\SU(2)$ throughout the Hurwitz orbit, so their pair traces lie in $[-2,2]^3$. Its real closure is compact and invariant under the chosen braid; for rational $c$ and a pseudo-Anosov iterate its complex Zariski closure is the Fricke surface by Corollary~\ref{cor:allbraids}. The gradient of the defining cubic at the initial trace triple is $(0,0,(a-2)(a-b))\ne0$ since $b<a<2$. Every orbit point is therefore smooth because each braid acts by an automorphism of the cubic. On the smooth chart $2z+xy-\kappa_{\rm F}\ne0$, the trace maps preserve the residue form $dx\wedge dy/(2z+xy-\kappa_{\rm F})$ \cite[equation~(1.6)]{CantatLoray}. Smoothness here concerns individual orbit points. Zariski density is compatible with a compact real orbit closure because the word-length lower bound measures growth at a finite place.

\begin{prop}[Trace field, signatures and an unbounded conjugate]\label{prop:arithmeticstructure}
For rational $c=A_c/B_c$, set $E_q=\Q(q)$, $F=\Q(q+q^{-1})$, and $\Gamma=\langle M_1,M_2,M_3\rangle$. Then $\Gamma\subset GL_3(\cO_{E_q}[1/B_c])$ preserves $H_{\rm q}$ and has adjoint trace field $F$. At $\sigma_k(q)=q^k$, $(k,N)=1$, the eigenvalues of its invariant form are
\[
 1,\qquad 1\pm\sqrt{1+2c^2}\,|\tan(\pi k/N)|.
\]
At $k_0=(N-1)/2$ the signature is $(2,1)$ and $\sigma_{k_0}(\Gamma)$ is unbounded. Every nonzero invariant $E_q$-Hermitian form is an $F^\times$-multiple of $H_{\rm q}$. In contrast to the unbounded output conjugate, every complex embedding of the input coefficient field $\Q(q,\sqrt{1-c^2})$ gives a unitary input tuple.
\end{prop}
\begin{proof}
The entries of each $M_j$ are integral away from $B_c$, and its determinant $-q$ is a unit. For $g\in\Gamma$, $\tr(\operatorname{Ad}_{\mathfrak{sl}_3}g)=\tr(g)\tr(g^{-1})-1=|\tr(g)|^2-1\in F$. Each generator gives the value $4-2(q+q^{-1})$, which generates $F$. These traces are invariant under scalars and conjugation, so the adjoint representation cannot descend to a proper subfield of $F$. Replacing $q$ by $q^k$ in the characteristic polynomial of $H_{\rm q}$ yields the displayed spectrum. At $k_0$ one has $|\tan(\pi k_0/N)|=\cot(\pi/(2N))>1$. At every embedding, $H^{[k]}=\sigma_k(H_{\rm q})$ is nonsingular. Given $w\ne0$ in an invariant subspace, some $e_j^*H^{[k]}w\ne0$. Applying $\sigma_k(M_j)-I=-(1+q^k)e_j e_j^*H^{[k]}$ puts $e_j$ in the subspace, and the nonzero off-diagonal entries of $H^{[k]}$ then give every $e_i$. Thus every conjugate is irreducible. If $J$ is another invariant form at the chosen embedding, $H_{\rm q}^{-1}J$ commutes with every generator. Schur's lemma gives $J=dH_{\rm q}$, with $d=J_{11}\in F^\times$ for nonzero $J$. Thus a change of basis or invariant form cannot make every embedding definite.

On $E_{13}=\langle e_1,e_3\rangle$, multiplication of the two blocks from \eqref{eq:familyM} gives the commutator $W=[M_1,M_3]$ determinant one and trace $2+2b(b-2)$. At $k_0$, $b$ becomes $-2\cos(\pi/N)<-1$. The conjugate block has an eigenvalue greater than one, so its powers are unbounded. For the input in \eqref{eq:familyT}, an embedding sends $q$ to a root of unity and $\sqrt{1-c^2}$ to one of its two real values. The normals remain real unit vectors, proving the assertion about all input embeddings.
\end{proof}
For $1\le k<N/2$ with $\gcd(k,N)=1$, the quotient-form formula of \cite[Theorem~1.1 and equation~(3)]{NegamiHermitian}, evaluated on the section of $\Pi$ with $\sigma_k(v_j)$ in slot $j$, gives $2\sin(2\pi k/N)H_{\rm q}^{[k]}$, where $H_{\rm q}^{[k]}=\sigma_k(H_{\rm q})$. The positive scalar preserves the signature above. More explicitly, let $\beta_1,\beta_2\in[0,1)$ be the eigenangles of $-\sigma_k(P_{\rm in})$, so its eigenvalues are $e^{2\pi i\beta_\nu}$. The input eigenangles are $0,k/N$ for each generator, so their three one-dimensional $1$-eigenspaces contribute $\dim K=3$ to the quotient construction. Also $\ker(-\sigma_k(P_{\rm in})-I)=0$ by Proposition~\ref{prop:unitary} and base change. The signature formula therefore gives negative index $1+3k/N-\beta_1-\beta_2$, equal to zero at $k=1$ and one at $k=k_0$.

In the parabolic cohomology setting, Dettweiler and Wewers state a signature formula for positive definite local systems without global sections \cite[Theorem~2.3]{DettweilerWewers}. Dettweiler and Sabbah describe how middle convolution changes Hodge data \cite{DettweilerSabbah}.

For comparison with the finite-image criterion of \cite[Theorem~2.10 and Corollary~2.11]{NegamiFinite}, a finite image would preserve a positive definite form at every embedding by averaging. At $k_0$, irreducibility makes any invariant Hermitian form a real multiple of the indefinite form just computed. This gives a second proof of infinitude, independent of the commutator calculation. The converse criterion in the cited work uses finite input and an invariant lattice. Our input has infinite image, and here the output matrices are integral away from the primes dividing $B_c$. The indefinite conjugate form precludes a totally definite model of the type in \cite[Section~3.1]{EvraParzanchevski}. Evra and Parzanchevski also study the index in the ambient arithmetic group for qutrit Clifford+$T$ \cite{EvraParzanchevskiThin}. The corresponding index for our reflection group remains undetermined.

\begin{corollary}[Exact finite-place contributions to the height]\label{cor:height}
Write $h(z)=\sum_v [E_v:\Q_v]\log\max(1,|z|_v)/[E:\Q]$ for the absolute logarithmic Weil height in a number field $E$ containing $z$ (Definition~\ref{def:heights}). For rational $c=A_c/B_c$ in lowest terms put $C=\log B_c-\log\gcd(B_c,\Phi_N(1))/\varphi(N)>0$. For every braid $\beta$,
\[
 2C|r_\beta+t_\beta|\ \le h(z_\beta)\le\ 2C|r_\beta+t_\beta|+\log 2 .
\]
The same bounds hold for $x_\beta,y_\beta$ with weights $|r_\beta|,|t_\beta|$. For $z_n$ along $LR$, the weight is $F_{2n}$ for $n\ge1$.
\end{corollary}
\begin{proof}
Compute the absolute logarithmic Weil height in $E_q$ (Definition~\ref{def:heights}). The coordinates are integral away from $B_c$. At each $p\mid B_c$, the local degrees over $p$ divided by $[E_q:\Q]$ sum to one, and Lemma~\ref{lem:familyval} gives the same $v(q-1)$ at every place over $p$. Corollary~\ref{cor:globalvaluations} gives a finite contribution of $2|r_\beta+t_\beta|\sum_{p\mid B_c}(\nu_p-v(q-1))\log p$. The sum equals $C$ since $\Phi_N(1)=p$ for $N$ a power of $p$ and equals one otherwise. Thus $C\ge(1-1/\varphi(N))\log B_c>0$. Every complex embedding preserves the reality and unit norms of the input normals, so each normalized pair trace has absolute value at most two. The archimedean contribution lies in $[0,\log2]$. The other coordinates follow identically, and \eqref{eq:fibov} gives the Fibonacci specialization.
\end{proof}
For $N=13$, $c=3/5$, one gets $C=\log5$ and $a=(32+9b)/25$; the bound at this point gives $2F_{2n}\log5\le h(z_n)\le 2F_{2n}\log5+\log2$ along $LR$ for $n\ge1$. For any pseudo-Anosov braid the $n$th roots of all three heights along its iterates tend to its stretch factor by the same slope calculation as in Corollary~\ref{cor:allbraids}.

\subsection{Mirrors and exact circuit counts}\label{sec:geometry}
Choose $C_{\rm q}^*C_{\rm q}=H_{\rm q}$. The columns $\psi_j=C_{\rm q}e_j$ have norm one. Put $\vartheta=\pi+2\pi/N$ and $\mathcal R_u(\theta)=I+(e^{i\theta}-1)uu^*$ for any unit vector $u$. Proposition~\ref{prop:unitary} gives $\widehat M_j=C_{\rm q}M_jC_{\rm q}^{-1}=\mathcal R_{\psi_j}(\vartheta)$. Its mirror is the fixed hyperplane $\psi_j^\perp$. Unitary conjugation carries the normal to its image, so $L,R$ act on a triple of normals $(u,v,w)$ of gates $(A,B,C)$ by $(Av,u,w)$ and $(u,w,C^{-1}v)$, respectively. Expanding two projectors gives
\[
 \frac{\tr(\widehat M_i\widehat M_j)-1}{-q}
   =2-(2+q+q^{-1})|\psi_i^*\psi_j|^2.
\]
Thus each transported pair trace records a squared overlap. Initially the squared overlaps in order $12,23,13$ are $\tan^2(\pi/N)(c^2,c^2,1)$ by the entries of $H_{\rm q}$.

\paragraph{Two-level implementation.}
Let $G_{ij}(\theta,\chi)$ act on levels $i,j$ by
$\left(\begin{smallmatrix}\cos\theta&-e^{-i\chi}\sin\theta\\e^{i\chi}\sin\theta&\cos\theta\end{smallmatrix}\right)$ and fix the other level. Put $D=\operatorname{diag}(-q,1,1)$, $\xi=\tan(\pi/N)$, $\rho=\sqrt{1-c^2\xi^2}$, and $\nu=\sqrt{1-(1+2c^2)\xi^2}/\rho$. A common Cholesky factor and alignments are
\[
 C_{\rm q}=\begin{pmatrix}1&-ic\xi&-i\xi\\0&\rho&-c\xi(\xi+i)/\rho\\0&0&\nu\end{pmatrix},
 \qquad
 \begin{aligned}
 R_1&=I_3,\qquad R_2=G_{01}(\arccos(c\xi),\pi/2),\\
 R_3&=G_{01}(\theta_a,-\pi/N)G_{02}(\theta_b,\pi/2),
 \end{aligned}
\]
where $\tan\theta_a=c\sqrt{1+\xi^2}/\rho$ and $\cos\theta_b=\xi\sqrt{1+c^2}/\rho$ with both angles in $(0,\pi/2)$. Direct multiplication gives $R_1|0\rangle=\psi_1$ and $R_j|0\rangle=i\psi_j$ for $j=2,3$. Hence $\widehat M_j=R_j D R_j^*$ uses at most $1,3,5$ elementary operations, counting each two-level rotation or one-level phase once. In this common basis each circuit first aligns the normal with $|0\rangle$, applies the phase, and undoes the alignment. On its two active levels $G_{ij}(\theta,\chi)=R_z(\chi)R_y(2\theta)R_z(-\chi)$ in the usual qubit rotation convention. Since $N$ is odd, $\widehat M_j^N=I_3-2\psi_j\psi_j^*$. Toffoli is the analogous reflection $I_8-2vv^*$ for $v=|11\rangle\otimes(|0\rangle-|1\rangle)/\sqrt2$: it swaps $|110\rangle$ and $|111\rangle$ and fixes the other basis states \cite{BarencoEtAl,AmyRossWesley}. Indeed $I_4\otimes H_{\rm Had}$ carries $|111\rangle$ to $v$, so conjugating the controlled-controlled-$Z$ by this target Hadamard gives Toffoli. The qutrit powers and Toffoli thus share the geometry of a phase flip on one line, in dimensions three and eight respectively.

For comparison, \cite[Appendix~A.3.1]{NegamiFinite} identifies the finite rank-two images $G_8$ and $G_{16}$ and gives their generator matrices. For $G_8$, take $-\lambda=i$ in those matrices $U,V$. Both have eigenvalues $1,i$, and $\tr(UV)=i$. In a basis orthonormal for the invariant Hermitian inner product, diagonalize $U$ with eigenvalues $1,i$ in that order. The traces then force both diagonal entries of $V$ to be $(1+i)/2$. Adjusting the relative basis phase therefore gives $S=\operatorname{diag}(1,i)$ and $R_x(\pi/2)$ up to phase. For $G_{16}$, choosing $-\lambda=e^{4\pi i/5}$ gives a generator unitarily conjugate to $\operatorname{diag}(1,e^{4\pi i/5})$ for the invariant inner product. Proposition~\ref{prop:density} establishes density for the present alphabet.

\paragraph{A fixed binary encoding.}
Encode the three levels as $|00\rangle,|01\rangle,|10\rangle$ and set $\mathsf E(U)=U\oplus 1$. All three levels carry input data here. In the generalized Toffoli circuits of Gokhale et al., the third level instead provides temporary workspace \cite{GokhaleQutrits}. Count CNOTs up to one global phase on all four levels, allowing arbitrary one-qubit unitaries at zero cost and no ancillas or measurements. This encoding depends on the chosen qutrit representative: $\mathsf E(I_3)$ costs zero, while $\mathsf E(-I_3)$ is controlled-$Z$ up to phase and costs one.
\begin{prop}[CNOT counts for encoded reflections]\label{prop:binarycost}
For a unit vector $\Psi\in\C^4$ and $\theta\not\equiv0\bmod 2\pi$, $\mathcal R_\Psi(\theta)$ requires three CNOTs if $\Psi$ is entangled. For a product vector it requires one CNOT when $\theta\equiv\pi\bmod 2\pi$ and two otherwise. The three $\mathsf E(\widehat M_j)$ therefore cost $(2,2,3)$ and their $N$th powers cost $(1,1,3)$ in the frame above.
\end{prop}
\begin{proof}
Write $Y=\left(\begin{smallmatrix}0&-i\\i&0\end{smallmatrix}\right)$, $J=Y\otimes Y$, $d=|\Psi^TJ\Psi|$, and $V=e^{-i\theta/4}\mathcal R_\Psi(\theta)$. Expanding gives
\[
 \tr(VJV^TJ)=4\cos(\theta/2)-4e^{i\theta/2}\sin^2(\theta/2)d^2.
\]
For an entangled vector $d=2|\Psi_{00}\Psi_{11}-\Psi_{01}\Psi_{10}|>0$, so the trace is nonreal. This excludes at most two CNOTs by \cite[Proposition~III.3]{ShendeBullockMarkov}; three suffice by \cite[Theorem~5]{VatanWilliams}. For a product vector local unitaries give a controlled phase, which entangles some product states and uses at most two CNOTs \cite[Corollary~5.3]{BarencoEtAl}. One CNOT requires zero trace \cite[Proposition~III.2]{ShendeBullockMarkov}, equivalent here to $\theta\equiv\pi$, when controlled-$Z$ attains the bound. The first two extended columns of $C_{\rm q}$ are product vectors and the third has $d=2c\xi\sqrt{1+\xi^2}\nu/\rho>0$. Substitution gives both triples.
\end{proof}
Every Hurwitz-transformed generator is still a reflection of phase $\vartheta$, hence costs two or three CNOTs in this encoding. Thus the sum for any pair stays between four and six CNOTs along the entire orbit, even when its minimum length in the fixed reflection alphabet grows exponentially by Theorem~\ref{thm:integrated}. Arbitrary exact one-qubit rotations are free in the CNOT count.

\section{Universal reflection gate sets in arbitrary rank}\label{sec:highrank}
We extend the cyclotomic construction to every dimension $r\ge3$. The targets are still generated by three-strand braid dynamics, while their synthesis may use all $r$ generators of a universal gate set. The additional generators need not preserve the three-dimensional subspace supporting the targets, so the lower bounds must be proved for the enlarged alphabet.

\begin{definition}[The rank-$r$ cyclotomic alphabet]\label{def:highrank}
Fix $r\ge3$, an odd integer $N\ge7$, and a real $0<c<1$, subject to
\begin{equation}\label{eq:highrankpositive}
 (r-1)\tan(\pi/N)<1.
\end{equation}
Put $q=e^{2\pi i/N}$, $\alpha=(1-q)/(1+q)$, and choose input normals in $\C^2$
\[
 v_2=(c,\sqrt{1-c^2})^T,\qquad v_j=(1,0)^T\quad(j\ne2),
 \qquad g_j=I_2+(q-1)v_jv_j^*.
\]
Write $b_{ij}=v_i^*v_j$: this is $c$ if exactly one of $i,j$ equals $2$, and $1$ otherwise. Define $H^{(r)}$ and $M_j^{(r)}$ by
\begin{equation}\label{eq:highrankmatrices}
 H^{(r)}_{ik}=\begin{cases}1,&i=k,\\
 \alpha b_{ik},&i<k,\\ \overline\alpha b_{ik},&i>k,
 \end{cases}
 \qquad
 M_j^{(r)}=I_r-(q+1)e_j(e_j^*H^{(r)}).
\end{equation}
Thus $M_j^{(r)}$ is the identity outside row $j$, whose entries are $-(q-1)b_{jk}$ for $k<j$, $-q$ for $k=j$, and $(q-1)b_{jk}$ for $k>j$. Denote the resulting alphabet by $\boldsymbol M^{(r)}$.
For $\beta\in B_3$, act on its first three entries and leave the other entries fixed; write $M_{j,\beta}^{(r)}$ for the resulting targets. Their costs are
\[
 \ell_r([V])=\len_{\boldsymbol M^{(r)}}([V]),\qquad
 L^{ij}_{r,\beta}(n)=\ell_r([M_{i,\beta^n}^{(r)}])+\ell_r([M_{j,\beta^n}^{(r)}]),
 \qquad
 D_{r,\beta}(n)=\sum_{j=1}^r\ell_r([M_{j,\beta^n}^{(r)}]).
\]
The symmetric alphabet defining $\ell_r$ contains every $(M_j^{(r)})^{\pm1}$, including $j>3$.
\end{definition}

\begin{theorem}[Universality and word costs in every rank]\label{thm:highrank}
For Definition~\ref{def:highrank}, the following assertions hold.
\begin{enumerate}[label=\textup{(\roman*)},leftmargin=*]
\item The tuple $\boldsymbol M^{(r)}$ is $\MC_{-1}(g_1,\ldots,g_r)$ on a quotient of dimension $r$. It preserves $H^{(r)}>0$. After one common unitarization, its generators are complex reflections of ordinary and projective order $2N$, and the generated projective group is dense in $\PU(r)$.
\item Suppose $c=A_c/B_c\in(0,1)$ is rational in lowest terms. For every $\beta\in B_3$, let $(r_\beta,t_\beta)$ be its primitive slope vector from Definition~\ref{def:symmetric-seed}. Then
\begin{equation}\label{eq:highranklower}
 L^{12}_{r,\beta}(1)\ge2|r_\beta|,\qquad
 L^{23}_{r,\beta}(1)\ge2|t_\beta|,\qquad
 L^{13}_{r,\beta}(1)\ge2|r_\beta+t_\beta|.
\end{equation}
For every pseudo-Anosov $\beta\in B_3$ and $1\le i<j\le3$,
\begin{equation}\label{eq:highrankgrowth}
 \lim_{n\to\infty}L^{ij}_{r,\beta}(n)^{1/n}
 =\lim_{n\to\infty}D_{r,\beta}(n)^{1/n}=\lambda_\beta.
\end{equation}
For $\phi=\LL\RR$ and every $n\ge1$,
\begin{equation}\label{eq:highrankfibonacci}
 2F_{2n}\le L^{13}_{r,\phi}(n)\le2F_{2n+2}.
\end{equation}
The constants in \eqref{eq:highranklower} and \eqref{eq:highrankfibonacci} are independent of $r,N,c$ subject to the stated hypotheses.
\end{enumerate}
\end{theorem}

\begin{proof}
\emph{Positive form and quotient realization.}
Since $\alpha=-i\tan(\pi/N)$ and $b_{ij}=b_{ji}$ is real, $H^{(r)}$ is Hermitian. Its diagonal entries are one, and in each row the sum of the absolute values of the off-diagonal entries is at most $(r-1)\tan(\pi/N)$. Hence
\[
 H^{(r)}\ \ge\bigl(1-(r-1)\tan(\pi/N)\bigr)I_r>0.
\]
The operator $P_j=e_j(e_j^*H^{(r)})$ is the $H^{(r)}$-orthogonal projection onto $\C e_j$, because $H^{(r)}_{jj}=1$. Thus $M_j^{(r)}=I_r-(q+1)P_j$ is $H^{(r)}$-unitary, with eigenvalues $-q$ and $1$ of multiplicities one and $r-1$. Its ordinary and projective orders are $2N$.

Use the $r$-block version of Definition~\ref{def:mcoperators}, with $\lambda=-1$, and put
\[
 \Pi(w_1,\ldots,w_r)=(v_1^*w_1,\ldots,v_r^*w_r),\qquad
 K=\bigoplus_{j=1}^r\ker(v_j^*).
\]
The map $\Pi$ is surjective with kernel $K$, and the row formulas give $\Pi G_j=M_j^{(r)}\Pi$. If $w\in L(-1)=\bigcap_j\ker(G_j-I)$, then $z=\Pi w$ is fixed by every $M_j^{(r)}$. Consequently $H^{(r)}z=0$, so $z=0$ and $w\in K$. On $K$, $G_j$ negates the $j$th summand and fixes the others, forcing $w=0$. Therefore $L(-1)=0$, and the quotient by $K$ is the middle-convolution space of dimension $r$. This directly verifies the quotient required by the general dimension formula \cite[Theorem~2.4(i)]{DRMiddle}.

\emph{Projective density.}
Choose $C_r^*C_r=H^{(r)}$ and put $\psi_j=C_re_j$. These unit vectors form a basis, with
$0<|\psi_i^*\psi_j|<1$ for $i\ne j$. The conjugated gates are
$I_r-(q+1)\psi_j\psi_j^*$. For every $j>1$, the first and $j$th gates preserve
$E_{1j}=\operatorname{span}(\psi_1,\psi_j)$ and fix its orthogonal complement pointwise. Their restrictions have projective order $2N>5$ and distinct rotation axes in $\PU(2)$. The closed-subgroup argument in Proposition~\ref{prop:density} gives projective density on $E_{1j}$. Since the determinants of these restrictions lie in the finite group generated by $-q$, their compact closure has identity component $\SU(E_{1j})$.

These plane groups generate $\SU(r)$. To see this explicitly, let
$V_k=\operatorname{span}(\psi_1,\ldots,\psi_k)$ and suppose the Lie algebra already contains $\mathfrak{su}(V_k)$, starting with $k=2$. Take $u=\psi_1$, a unit vector $e\in V_{k+1}\cap V_k^\perp$, and write
$\psi_{k+1}=a u+v+b e$, where $v\in V_k\cap u^\perp$ and $b>0$ after choosing the phase of $e$. The algebra $\mathfrak{su}(E_{1,k+1})$ contains
\[
 u(v+be)^*-(v+be)u^*,\qquad
 i\bigl(u(v+be)^*+(v+be)u^*\bigr).
\]
Subtract the terms involving $v$, which lie in $\mathfrak{su}(V_k)$. The resulting two couplings between $u$ and $e$, together with their brackets and $\mathfrak{su}(V_k)$, generate $\mathfrak{su}(V_{k+1})$. Induction proves that the full compact closure contains $\SU(r)$, and hence its projective image is $\PU(r)$.

\emph{Traces and the cost of all generators.}
Let $E_3=\operatorname{span}(e_1,e_2,e_3)$. The first three generators preserve $E_3$, restrict there to the original qutrit matrices \eqref{eq:familyM}, and fix $E_3^{\perp_{H^{(r)}}}$ pointwise. The same is true after every three-strand Hurwitz move. If $t_{ij,\beta}$ is respectively $x_\beta,y_\beta,z_\beta$ for $ij=12,23,13$, Lemma~\ref{lem:transport} therefore gives
\begin{equation}\label{eq:highranktrace}
 \tr\bigl(M_{i,\beta}^{(r)}M_{j,\beta}^{(r)}\bigr)
 =(r-3)+\tr(M_{i,\beta}M_{j,\beta})=r-2-q\,t_{ij,\beta}.
\end{equation}
This trace identity concerns the targets; it does not restrict the generators allowed in their synthesis.

Choose a prime $p\mid B_c$, a place of $\Q(q)$ above $p$ normalized by $v(p)=1$, and put
$\delta=\operatorname{ord}_p(B_c)-v(q-1)>0$ as in Lemma~\ref{lem:familyval}. Every entry of every $M_j^{(r)}$ has valuation at least $-\delta$: its off-diagonal entries are signed copies of $q-1$ or $c(q-1)$. The same bound holds for every inverse, since
\[
 (M_j^{(r)})^{-1}=q^{-1}M_j^{(r)}+(1-q^{-1})I_r.
\]
All determinants equal $-q$ and are valuation units. If $u$ denotes the corresponding positive slope weight $|r_\beta|$, $|t_\beta|$, or $|r_\beta+t_\beta|$, Corollary~\ref{cor:globalvaluations} gives
$v(t_{ij,\beta})=-2\delta u$. In \eqref{eq:highranktrace}, the integer $r-2$ has nonnegative valuation and the other term has negative valuation. Thus the full rank-$r$ pair trace has valuation $-2\delta u$. Applying Lemma~\ref{lem:denominator} to the entire rank-$r$ alphabet gives $\delta L^{ij}_{r,\beta}(1)\ge2\delta u$. If $u=0$, the bound is trivial. This proves \eqref{eq:highranklower}, including optimization over all additional gates.

For pseudo-Anosov $\beta$, the three slope weights along $\beta^n$ have $n$th-root limit $\lambda_\beta$, as in Corollary~\ref{cor:allbraids}. Their lower bounds and the three-generator representing words of Lemma~\ref{lem:upper} prove the pair limits. The remaining $r-3$ targets are unchanged and have length one, so
\[
 L^{ij}_{r,\beta}(n)\le D_{r,\beta}(n)
 \le3\mathfrak b_n(\beta)+(r-3),
\]
which proves the total-cost limit. Finally, the Fibonacci trace valuations and the explicit representing words from Theorem~\ref{thm:fibonacci} give \eqref{eq:highrankfibonacci}.
\end{proof}

For each fixed $r$, condition~\eqref{eq:highrankpositive} holds for all sufficiently large odd $N$; it is a sufficient positivity condition, not a claimed sharp threshold. The whole alphabet is universal on $\C^r$, although the targets used for the lower bounds act on a three-dimensional subspace. The theorem concerns these embedded $B_3$ dynamics and fixed-alphabet costs; it does not assert the exact shortest-length equalities of Theorem~\ref{thm:sharpwords} for the enlarged alphabet, or an efficient implementation by local multiqubit gates.

For periodic and non-periodic reducible $\beta\in B_3$, Lemma~\ref{lem:nonpa} gives $L^{ij}_{r,\beta}(n),D_{r,\beta}(n)=O_\beta(n+1)$, with $r$ fixed. Every target is a non-scalar reflection, so $L^{ij}_{r,\beta}(n)\ge2$ and the $n$th-root limits of both costs are $1$. No linear lower bound is asserted for every such braid and pair.

\section{Generic faithfulness and exact finite-time lengths}\label{sec:geodesicity}
We prove Theorem~\ref{thm:sharpwords}. Throughout this section $N\ge7$ is odd,
$q=e^{2\pi i/N}$, and $E_q=\Q(q)$. Let $X$ be an indeterminate and write $M_j(X)$ for the matrices in \eqref{eq:familyM} over $E_q(X)$. Specializing $X=c$ gives $M_j(c)$. A geodesic word means a shortest
word in the indicated symmetric generating set, with every letter of cost one. The faithfulness proof detects a mixed word by a nonzero rank-one leading coefficient in $t=X^2$. The proof of minimality then shows that the triangle-subgroup blocks in the recursive words have length at most five and are geodesic.

\begin{definition}[Projective gate groups with specified generators]\label{def:markedgroups}
For a fixed parameter $c\in\C$, let the specialized matrices in \eqref{eq:familyM} generate the projective group
\[
\Gamma_{N,c}=\langle[M_1(c)],[M_2(c)],[M_3(c)]\rangle.
\]
Define
\begin{equation}\label{eq:abstractH}
\mathcal H_N=\langle\mathsf a,\mathsf h,\mathsf z\mid
\mathsf a^{2N}=\mathsf h^{2N}=\mathsf z^2=1,\
(\mathsf a\mathsf h)^N=\mathsf z,\
[\mathsf z,\mathsf a]=[\mathsf z,\mathsf h]=1\rangle.
\end{equation}
and put $\mathcal F_N=\mathcal H_N*\langle\mathsf b\mid\mathsf b^{2N}=1\rangle$.
\end{definition}
Here $C_m$ denotes the cyclic group of order $m$ and $\Delta(l,m,n)=\langle u,v\mid u^l=v^m=(uv)^n=1\rangle$ the orientation-preserving triangle group. The group $\mathcal H_N$ is a central $C_2$-extension of the orientation-preserving triangle group $\Delta(2N,2N,N)$; Proposition~\ref{prop:trianglesubgroup} identifies its concrete two-generator realization. The complete shortest-word statement is as follows.

\begin{maintheorem}[Generic faithfulness and sharp finite-time costs]\label{thm:sharpwords}
For every odd $N\ge7$, the following hold.
\begin{enumerate}[label=\textup{(\roman*)},leftmargin=*]
\item The assignments $\mathsf a\mapsto[M_1(X)]$, $\mathsf b\mapsto[M_2(X)]$, and $\mathsf h\mapsto[M_3(X)]$ define a faithful representation $\mathcal F_N\to PGL_3(E_q(X))$.
For every transcendental real $c\in(0,1)$, evaluation at $X=c$ gives an isomorphism
$\mathcal F_N\simeq\Gamma_{N,c}$.
\item At every such transcendental parameter, the individual Fibonacci
targets have exact lengths, for all $n\ge1$,
\begin{equation}\label{eq:exactindividual}
\begin{aligned}
\len_{\boldsymbol M}([M_{1,n}])&=2F_{2n+1}-1,&
\len_{\boldsymbol M}([M_{2,n}])&=2F_{2n-1}-1,\\
\len_{\boldsymbol M}([M_{3,n}])&=2F_{2n}+1,&
L^{13}_{\boldsymbol M,\phi}(n)&=2F_{2n+2},
\end{aligned}
\end{equation}
where $\boldsymbol M_n=\phi^n\boldsymbol M$ and $\phi=\LL\RR$.
\item Fix a horizon $m_\ast\ge1$ and put $R_{m_\ast}=2F_{2m_\ast+1}-1$.
For every reduced rational $c=A_c/B_c\in(0,1)$ satisfying
\begin{equation}\label{eq:finitehorizoncondition}
B_c^2>2\cdot5^{\,2R_{m_\ast}-1},
\end{equation}
all equalities in \eqref{eq:exactindividual} hold for $1\le n\le m_\ast$.
In particular $B_c\ge5^{R_{m_\ast}}$ suffices, uniformly in $N$ and $A_c$.
\end{enumerate}
\end{maintheorem}

\begin{remark}[An algebraic specialization with a new relation]
Put $c_*=1/(2\cos(\pi/(2N)))$. In \eqref{eq:familyM} one has
\[
 \det(zI-M_1M_2)=(z-1)(z^2+qa z+q^2),\qquad a=2-4c^2\sin^2(\pi/N).
\]
At $c_*$, $a=2\cos(\pi/N)$ and the three eigenvalues $1,-qe^{i\pi/N},-qe^{-i\pi/N}$ are distinct $N$th roots of unity since $N$ is odd. Thus $(M_1M_2)^N=I$, whereas $(\mathsf a\mathsf b)^N$ is reduced and nontrivial in the free product $\mathcal F_N$. Generic faithfulness can therefore fail at an algebraic parameter.
\end{remark}

\subsection{A triangle subgroup and its short words}
\begin{prop}[The two-generator subgroup]\label{prop:trianglesubgroup}
Evaluating the polynomial matrices \eqref{eq:familyM} at any $c\in\C$, the map $\mathsf a\mapsto M_1(c)$,
$\mathsf h\mapsto M_3(c)$ identifies the group $\mathcal H_N$ of
\eqref{eq:abstractH} with $\langle M_1(c),M_3(c)\rangle$.
This subgroup maps injectively to $PGL_3$, and fits into
\[
1\longrightarrow\langle\mathsf z\rangle\simeq C_2
\longrightarrow\mathcal H_N\longrightarrow\Delta(2N,2N,N)
\longrightarrow1.
\]
Here $\Delta$ denotes the orientation-preserving projective triangle group.
Every freely reduced word of length at most five in
$\mathsf a^{\pm1},\mathsf h^{\pm1}$ is geodesic in $\mathcal H_N$.
\end{prop}
\begin{proof}
Put $d_q=q-1$, $D_q=q^2+1$, and
$w_c=e_2+cd_q(e_1-qe_3)/D_q$. The basis $(e_1,e_3,w_c)$ sends $M_1,M_3$ to
$\operatorname{diag}(U,1),\operatorname{diag}(V,1)$, where
\begin{equation}\label{eq:triangleblocks}
U=\begin{pmatrix}-q&d_q\\0&1\end{pmatrix},\qquad
V=\begin{pmatrix}1&0\\-d_q&-q\end{pmatrix}.
\end{equation}
Thus the marked subgroup is independent of $c$, and its only scalar
$3\times3$ matrix is the identity. The eigenvalues of $UV$ are $-1,-q^2$,
so $(UV)^N=-I_2$, while $U^{2N}=V^{2N}=I_2$.

Under the automorphism $q\mapsto q^{(N-1)/2}=-e^{-\pi i/N}$, write
$U',V'$ for the blocks and put $\eta_N=e^{-\pi i/(2N)}$. The determinant-one
pair $U'/\eta_N,V'/\eta_N$ has traces
\[
x=y=2\cos\frac{\pi}{2N},\qquad
z=\tr(U'V'/\eta_N^2)=-2\cos\frac{\pi}{N}.
\]
To see the sign geometrically, put $\alpha=\pi/(2N)$. For the triangle of angles $\alpha,\alpha,2\alpha$, the side $d$ between the first two vertices satisfies $\cosh(d)\sin^2\alpha=\cos^2\alpha+\cos(2\alpha)$ by the hyperbolic cosine law. The positive-trace lifts of the two equally oriented vertex rotations thus have product trace $2(\cos^2\alpha-\cosh(d)\sin^2\alpha)=-2\cos(2\alpha)$, in the convention of \cite[Section~2]{McMullenTriangle}. These traces determine the pair up to
simultaneous conjugacy. Indeed, $x^2+y^2+z^2-xyz-2>2$ gives irreducibility;
after diagonalizing the first matrix, $y,z$ determine the diagonal entries
of the second, and its determinant determines the product of its nonzero
off-diagonal entries. A diagonal conjugation adjusts the latter entries.
It follows that the projective block image is faithfully
$\Delta(2N,2N,N)$. Field automorphisms preserve matrix relations.

Any word scalar on the two-dimensional block is therefore in the normal
closure of $\mathsf a^{2N},\mathsf h^{2N},(\mathsf a\mathsf h)^N$.
Their block values are $I_2,I_2,-I_2$, so the scalar kernel of the block
representation is exactly $\{\pm I_2\}$. The presentation
\eqref{eq:abstractH} follows: its quotient by $\mathsf z$ is the triangle
presentation, and $\mathsf z$ maps to $\operatorname{diag}(-I_2,1)\ne I_3$.
In particular this central involution survives three-dimensional
projectivization.

For the length assertion, apply small cancellation in the triangle quotient $\Delta(2N,2N,N)$. Take the set of all distinct cyclic shifts of
$\mathsf a^{2N},\mathsf h^{2N},(\mathsf a\mathsf h)^N$ and their inverses.
Any two distinct words in this symmetrized set have common initial
segments of length at most one. Since all relators have length $2N\ge14$
and $1<2N/12$, the presentation satisfies $C'(1/12)$.
Definition~3.6 of \cite{OOS} uses distinct words to define pieces and
does not exclude proper powers. Greendlinger's lemma in the form of
\cite[Lemma~3.7]{OOS}, with parameter $1/12\le1/6$, implies that a
nonempty cyclically reduced word representing the identity contains a relator segment of
length greater than $3N/2\ge10.5$. Cyclic reduction therefore rules out
every nonempty freely reduced word of length at most ten representing the identity.
A reduced word of length at most five and a shorter representative
would give such a word representing the identity, of length at most nine. Hence short words
are geodesic in the triangle quotient and, consequently, in $\mathcal H_N$.
\end{proof}

\subsection{Nonvanishing of mixed-word coefficients}
Work over $E_q(X)$. Substituting $X$ for $c$ in the preceding basis and scaling its first two vectors by $X$ gives $S(X)=(Xe_1,Xe_3,w_X)$. Put $t=X^2$. The first and third matrices remain
$\widehat U=\operatorname{diag}(U,1)$ and $\widehat V=\operatorname{diag}(V,1)$.
The second becomes
\begin{equation}\label{eq:mixedrankone}
\mathcal B(t)=S(X)^{-1}M_2(X)S(X)=I_3+u(\omega_0+t\omega_1),
\end{equation}
where
\[
u=\begin{pmatrix}-d_q/D_q\\qd_q/D_q\\1\end{pmatrix},\qquad
\omega_0=(0,0,-q-1),\qquad
\omega_1=\left(-d_q,d_q,-\frac{(q+1)d_q^2}{D_q}\right).
\]
Since $\omega_0u=-q-1$ and $\omega_1u=0$, for every integer $k$ we have
\begin{equation}\label{eq:mixedpowers}
\mathcal B(t)^k=I_3+s_k u(\omega_0+t\omega_1),\qquad
s_k=\frac{1-(-q)^k}{1+q}.
\end{equation}
The scalar $s_k$ is nonzero whenever $k\not\equiv0\pmod{2N}$.

\begin{lemma}[Nonzero internal coefficients]\label{lem:nonzeroconnector}
If $g=\operatorname{diag}(P,1)\in\mathcal H_N$ is nonidentity, then
$\omega_1gu\ne0$.
\end{lemma}
\begin{proof}
Let $v=(-1,q)^T$ and $f=(-1,1)$. Direct multiplication gives
\[
\omega_1gu=\frac{d_q^2}{D_q}f(P-I_2)v.
\]
The blocks $U,V$ preserve the Hermitian form
\[
J=\begin{pmatrix}1&(1-q)/(1+q)\\(q-1)/(q+1)&1\end{pmatrix}.
\]
Its eigenvalues are $1\pm\tan(\pi/N)>0$, and
\[
v^*J=\frac{D_q}{q(q+1)}f,\qquad v^*Jv=q+q^{-1}>0.
\]
Consequently $\omega_1gu=0$ would give $v^*JPv=v^*Jv$.
Since $P$ is $J$-unitary, expanding $(Pv-v)^*J(Pv-v)$ then gives
$Pv=v$.

We show that the linear stabilizer of $v$ is trivial. One has
$VUv=-q^2v$. After the same Galois embedding as in
Proposition~\ref{prop:trianglesubgroup}, $J$ has signature $(1,1)$ and
$v^*Jv=-2\cos(\pi/N)<0$. Thus $[v]$ lies in the disk of negative lines.
We first determine the stabilizer of the line $[v]$, then its subgroup fixing the vector $v$. The projective element $[VU]$ is conjugate to $[UV]$, the inverse of the
canonical third triangle generator. Its fixed point is the order-$N$
vertex. The doubled triangle is a fundamental domain with angle
$2\pi/N$ at this vertex, so its full stabilizer has order $N$ and is
$\langle[VU]\rangle$; see \cite[Section~2]{VoightTriangle} for the
vertex and stabilizer conventions. This uses only the geometric triangle tessellation.
Its inverse image in $\langle U,V\rangle$ is exactly $\langle VU\rangle$,
because $(VU)^N=-I_2$ and the scalar kernel is $\{\pm I_2\}$.
The eigenvalue $-q^2$ has order $2N$ for odd $N$, so no nonidentity
element of this inverse image fixes $v$ as a vector. Galois embeddings
preserve such vector equations, proving the assertion.
\end{proof}

\begin{proof}[Proof of Theorem~\ref{thm:sharpwords}(i)]
By Proposition~\ref{prop:trianglesubgroup} and \eqref{eq:mixedpowers}, the
matrices define the marked homomorphism from $\mathcal F_N$.
In free-product normal form each syllable is a nonidentity element of one factor, and successive syllables belong to different factors \cite[Chapter~IV]{LyndonSchupp}. With $m\ge1$ syllables from the cyclic factor, the word has the form
\[
h_0\mathcal B(t)^{k_1}h_1\cdots h_{m-1}\mathcal B(t)^{k_m}h_m,
\]
where the internal $h_j\in\mathcal H_N$ are nonidentity and
$k_j\not\equiv0\pmod{2N}$. Its coefficient of $t^m$ is
\begin{equation}\label{eq:mixedleading}
\left(\prod_{j=1}^m s_{k_j}\right)
\left(\prod_{j=1}^{m-1}\omega_1h_ju\right)(h_0u)(\omega_1h_m).
\end{equation}
The lemma makes this a nonzero rank-one matrix. The word therefore has
degree exactly $m$ and is not scalar: the nonzero leading coefficient
of a scalar $3\times3$ polynomial matrix has rank three. If $m=0$,
projective injectivity follows from Proposition~\ref{prop:trianglesubgroup}.
The representation is therefore faithful over $E_q(t)$. Extending scalars via $t\mapsto X^2$ and returning to the original basis gives faithfulness over $E_q(X)$. For every transcendental real $c\in(0,1)$, evaluation at $X=c$ embeds $E_q(X)$ in $\C$ and preserves faithfulness.
\end{proof}

\subsection{Geodesicity of the recursive words}
Let $W_{1,n},W_{2,n},W_{3,n}$ be the formal Hurwitz words starting at
$(\mathsf a,\mathsf b,\mathsf h)$ and satisfying
\[
(W_{1,n+1},W_{2,n+1},W_{3,n+1})
=(W_{1,n}W_{3,n}W_{1,n}^{-1},\ W_{1,n},\
W_{3,n}^{-1}W_{2,n}W_{3,n}).
\]
\begin{lemma}[Geodesic words in the abstract free product]\label{lem:freeproductgeodesics}
For every $n\ge1$, these words are geodesic in $\mathcal F_N$ for the
six letters $\mathsf a^{\pm1},\mathsf b^{\pm1},\mathsf h^{\pm1}$.
Their lengths are respectively $2F_{2n+1}-1$, $2F_{2n-1}-1$, and $2F_{2n}+1$.
\end{lemma}
\begin{proof}
A block is a maximal consecutive subword in $\mathsf a^{\pm1},\mathsf h^{\pm1}$.
Set
\[
W_{\rm com}=\mathsf a\mathsf h\mathsf a^{-1}\mathsf h^{-1},\qquad
\mathcal I=\{\mathsf h\mathsf a\mathsf h^{-1},\
\mathsf h\mathsf a^{-1}\mathsf h^{-1},\
\mathsf h\mathsf a\mathsf h^{-1}\mathsf a^{-1}\mathsf h^{-1},\
\mathsf h\mathsf a\mathsf h\mathsf a^{-1}\mathsf h^{-1}\}.
\]
The set $\mathcal I$ is closed under inversion. At $n=1$ the words are
$\mathsf a\mathsf h\mathsf a^{-1},\mathsf a,\mathsf h^{-1}\mathsf b\mathsf h$.
At $n=2$ the block lists are $(W_{\rm com},W_{\rm com}^{-1})$,
$(\mathsf a\mathsf h\mathsf a^{-1})$, and
$(\mathsf h^{-1},\mathsf h\mathsf a\mathsf h^{-1},\mathsf h)$.
For instance, the junction between $W_{1,2}$ and $W_{3,2}$ in $W_{1,3}$ is $W_{\rm com}^{-1}\mathsf h^{-1}=\mathsf h\mathsf a\mathsf h^{-1}\mathsf a^{-1}\mathsf h^{-1}$, a reduced internal block. The same substitution gives at $n=3$:
the boundary blocks of $W_{1,n},W_{2,n}$ are $W_{\rm com},W_{\rm com}^{-1}$, those of
$W_{3,n}$ are $\mathsf h^{-1},\mathsf h$, every internal block lies in
$\mathcal I$, and every intervening $\mathsf b$-syllable is one letter.
For induction, the only new merged blocks are
$W_{\rm com}^{-1}\mathsf h^{-1}$ and $\mathsf hW_{\rm com}$, the last two members of
$\mathcal I$. They are freely reduced and nonempty; all other internal
blocks are inherited or inverted. No adjacent $\mathsf b$-letters arise.
Thus every block has length at most five, for all $n$.

Proposition~\ref{prop:trianglesubgroup} makes all these blocks nonidentity
and geodesic in $\mathcal H_N$. Each $\mathsf b$-syllable is geodesic in
$C_{2N}$. The free-product normal-form theorem now gives geodesicity of
the whole words in the union generating set: reducing any competing word
to its unique alternating normal form cannot increase the sum of the
factor word lengths. There is no free cancellation,
so their lengths satisfy the recurrence in the proof of
Theorem~\ref{thm:fibonacci}, giving the stated formulas.
\end{proof}
Together with part (i), this proves Theorem~\ref{thm:sharpwords}(ii).

\subsection{Effective rational specialization}
\begin{prop}[Lower bound for a new relation]\label{prop:newrelationbound}
Suppose a word of written length $d_0$ in the six original letters
is nonidentity in $\mathcal F_N$ but becomes projectively trivial at a
reduced rational parameter $c=A_c/B_c\in(0,1)$. Then
\begin{equation}\label{eq:newrelationbound}
B_c^2\le2\cdot5^{d_0}.
\end{equation}
\end{prop}
\begin{proof}
All six original letters have entries in $\cO_{E_q}[X]=\mathbb Z[q][X]$ of
degree at most one, since $M_j^{-1}=I_3+(M_j-I_3)/q$.
For each embedding $\sigma:E_q\hookrightarrow\C$, the coefficient norm of a polynomial matrix $P$
\[
\|P\|_{\sigma,\mathrm{coef}}
=\max_i\sum_j\sum_{r\ge0}|\sigma([X^r]P_{ij})|
\]
is submultiplicative. Every letter has norm at most $5$, because
$|\sigma(q)|=1$ and $|\sigma(q-1)|\le2$. Each coefficient of every
entry of a length-$d_0$ word is therefore at most $5^{d_0}$ in modulus
at every embedding.

Generic faithfulness gives a nonzero witness polynomial $f(X)$, either
an off-diagonal entry or the difference of two diagonal entries.
If the word specializes to a scalar matrix at $X=c$, then $f(c)=0$, and
each coefficient has all its conjugates bounded by $2\cdot5^{d_0}$.
For $J_0=\operatorname{diag}(1,-1,1)$, every letter satisfies
$M_j(-X)^{\pm1}=J_0M_j(X)^{\pm1}J_0^{-1}$. Thus every word entry has
fixed parity, and diagonal differences are even. We may write
\[
f(X)=X^\epsilon Q(X^2),\qquad \epsilon\in\{0,1\},\quad
0\ne Q\in\cO_{E_q}[t].
\]
Since $c\ne0$, the equality $f(c)=0$ forces $Q(c^2)=0$. Thus $Q$ is nonconstant. Write
$d_Q=\deg Q\ge1$ and let $\alpha\ne0$ be its leading coefficient.
From $Q(A_c^2/B_c^2)=0$, multiplication by $B_c^{2d_Q}$ and reduction
modulo $B_c^2\cO_{E_q}$ give $\alpha A_c^{2d_Q}\in B_c^2\cO_{E_q}$.
The integers $A_c^{2d_Q}$ and $B_c^2$ are coprime, so
$\alpha\in B_c^2\cO_{E_q}$. For $e=[E_q:\Q]$, the field norm gives
\[
B_c^{2e}\le|N_{E_q/\Q}(\alpha)|\le(2\cdot5^{d_0})^e,
\]
proving the claim.
\end{proof}

\begin{proof}[Proof of Theorem~\ref{thm:sharpwords}(iii)]
Lemma~\ref{lem:freeproductgeodesics} gives a geodesic representative
of length $\ell\le R_{m_\ast}$ for each target through time $m_\ast$.
If a shorter word represented the same projective target after
specialization, their quotient would be nonidentity in $\mathcal F_N$,
projectively trivial after specialization, and of written length at
most $2\ell-1\le2R_{m_\ast}-1$. Proposition~\ref{prop:newrelationbound}
contradicts \eqref{eq:finitehorizoncondition}. The simpler sufficient
condition follows from $5^{2R_{m_\ast}}>2\cdot5^{2R_{m_\ast}-1}$.
\end{proof}

\begin{remark}[Quantifiers and the remaining gap]\label{rem:sharpquantifiers}
For each prescribed finite horizon, only finitely many reduced rational
parameters in $(0,1)$ fail the sufficient denominator condition.
For horizons $m_\ast=1,2,3$ the simpler thresholds are respectively
$5^3,5^9,5^{25}$, and the exact first--third pair costs through those
times are $6$; $6,16$; and $6,16,42$. The estimate is conservative and
failure of the condition does not imply a shortening. A large denominator
also does not require $c$ to be numerically small, since every coprime
numerator $0<A_c<B_c$ is allowed. The denominator threshold grows with
the horizon; the theorem does not give all-time geodesicity at a single
fixed rational parameter such as $c=3/5$. The transcendental choice $c=e^{-1}$ gives the same first three costs and satisfies the all-time formula. For fixed $N$, its gate amplitudes meet the polynomial-time computability condition of \cite[Definition~3.2.2 of the author version]{BernsteinVazirani}: the Cholesky formula in Subsection~\ref{sec:geometry} uses a fixed number of operations and positive square roots on constants computable to $p$ bits in time polynomial in $p$.
\end{remark}

\subsection{A fixed rational example}
\begin{prop}[Verification by reduction modulo 53]\label{prop:f53}
At $N=13$ and $c=3/5$, the individual projective lengths in the fixed six-letter alphabet are $(3,1,3)$ after one iteration of $LR$ and $(9,3,7)$ after two. The first and third lengths sum to $6$ and $16$.
\end{prop}
\begin{proof}
Reduce $\mathbb Z[q,1/5]$ to $\mathbb F_{53}$ by $q\mapsto16$ and $c\mapsto43$. The former has order $13$ and the generators have unit determinants at this place. Hence any scalar relating two representing words is a local unit, and projective equality survives reduction.

Breadth-first enumeration in $\operatorname{PGL}_3(\mathbb F_{53})$ gives $937$ classes in the radius-four ball $B$. Normalize each matrix by its first nonzero entry. For a target $T$, minimize $d(I,U)+d(I,U^{-1}T)$ over $U\in B$ with $U^{-1}T\in B$. This tests every word of length at most eight by splitting it in two. Amy et al. use the same meet-in-the-middle principle to search for depth-optimal circuits \cite{AmyMITM}. The resulting distances for the two iterates are $(3,1,3)$ and $(>8,3,7)$, respectively; \texttt{verify\_f53.py} reproduces the enumeration. The explicit Hurwitz words have lengths $(3,1,3)$ and $(9,3,7)$, supplying matching upper bounds in the original group.
\end{proof}
For this fixed rational point, equality with the Fibonacci upper bound for all $n$ remains open. Appendix~\ref{app:local} proves the all-time lower bound $2F_{2n}+2$ by following the poles of target prefixes at a place above five. Each nonempty block in the triangle generators that occurs in these words contributes one pole, even when the block has subgroup word length five. The pole count therefore loses information about length within a block.

\section{Exact word costs and approximation}\label{sec:scope}
Fix odd $N\ge7$ and rational $0<c<1$. Use the upper-triangular Cholesky factor $C_{\rm q}$ with positive diagonal from Subsection~\ref{sec:geometry}, and write $\mathcal G=\{(C_{\rm q}M_jC_{\rm q}^{-1})^{\pm1}:1\le j\le3\}$. Let $\ell_{\mathcal G,\epsilon}([U])$ be the minimum length of a word whose matrix is within operator-norm distance $\epsilon$ of $U$ up to phase. The worst-case approximate word length at accuracy $\epsilon$ is $\mathcal C_{\mathcal G}(\epsilon)=\sup_{[U]\in\PU(3)}\ell_{\mathcal G,\epsilon}([U])$.

\begin{corollary}[Optimal-order approximate word length]\label{cor:logapprox}
For each such alphabet, $\mathcal C_{\mathcal G}(\epsilon)=\Theta_{N,c}(\log(1/\epsilon))$ as $\epsilon\downarrow0$. The upper bound is uniform over all targets, and the lower bound is for the worst case.
\end{corollary}
\begin{proof}
The chosen factor has algebraic entries. Scale the three generators by an algebraic cube root of $(-q)^{-1}$ to give determinant-one lifts. Their compact closure projects onto $\PU(3)$ by Proposition~\ref{prop:density} and hence has the full Lie algebra of $\SU(3)$. The algebraic lifts therefore generate a dense group and have a spectral gap by \cite[Theorem~1]{BourgainGamburd}. For the symmetric walk made lazy by placing half the mass at the identity \cite[Chapter~1]{LevinPeres}, the averaging operator has nonnegative spectrum and norm strictly below one on mean-zero functions. The covering argument of \cite[Theorem~1, equations~(12)--(15)]{HarrowRechtChuang} gives words of length $O_{N,c}(\log(1/\epsilon))$ for every target. Identity steps can be dropped. A simultaneous unitary change of basis of the generators and targets preserves operator-norm distance and hence all minimum approximate lengths.

Conversely there are at most $(3\cdot5^L-1)/2$ freely reduced words of length at most $L$. An $\epsilon$-ball in $\PU(3)$ has Haar volume $O(\epsilon^8)$ since the group has dimension eight and the projective operator norm is locally comparable to a Riemannian distance. Covering the group requires $L\ge 8\log(1/\epsilon)/\log5-O(1)$.
\end{proof}

\begin{prop}[Exact length, approximate length and matrix evaluation]\label{prop:resourcecontrast}
For a fixed braid $\beta$, put $(U_{1,n},U_{2,n},U_{3,n})=\beta^n(\widehat M_1,\widehat M_2,\widehat M_3)$ in this Cholesky basis. For the fixed rational alphabet above, the sum of their three minimum approximate lengths is $O_{N,c}(1+\log(1/\epsilon))$ uniformly in $n$ for $0<\epsilon<1/2$. Their matrices can be computed classically to operator-norm error $\epsilon$ in time polynomial in the iteration count $n$ and in $\log(1/\epsilon)$. For the Fibonacci braid, the exact first--third pair cost is $\Theta(\Lambda^n)$.
\end{prop}
\begin{proof}
Sum the uniform upper bound of Corollary~\ref{cor:logapprox} over the three targets. For matrix evaluation, a fixed braid requires $O(n)$ updates on three $3\times3$ matrices. If the current maximum error is $E\le1$, the approximate factors have norm at most two. Telescoping a conjugation update using the adjoint bounds the new error by $(4+2+1)E+O(2^{-p})$ at $p$ bits. Thus $p=O(n+\log(1/\epsilon))$ suffices. The algebraic input and each fixed-size matrix update take polynomial time in $p$. The exact length estimate is Theorem~\ref{thm:fibonacci}.
\end{proof}
For a physical implementation by $L$ unitary factors, telescoping the product bounds its error by the sum of the individual operator-norm errors. Accuracy $\epsilon/L$ per factor therefore suffices. Exact minimality alone gives no quantitative separation from shorter approximate words. The gap argument here uses rational $c$, which makes the gates algebraic. Finding a short circuit is a separate task from evaluating its target matrix. Solovay--Kitaev compilation and its refinements provide algorithmic polylogarithmic bounds \cite{DawsonNielsen,Kuperberg}, while efficient navigation is a further requirement for golden gates \cite[Definition~2.8]{EvraParzanchevski}, as illustrated in $PU(2)$ by \cite{ParzanchevskiSarnak}.

\section{Discussion}\label{sec:discussion}
Householder and Givens synthesis allow the reflection directions or rotation parameters to depend on the target \cite{IvanovHouseholder,VitanovQutrit}. Our alphabet fixes three directions. Optimal qutrit Clifford+$T$ synthesis also uses a fixed alphabet, with the number of $T$ gates as cost \cite{Prakash,GRT}. Denominator reduction appears in exact single- and multiqubit Clifford+$T$ synthesis \cite{KliuchnikovMaslovMosca,GilesSelinger}. Related arithmetic approaches include \cite{ForestEtAl,KalraValluriMosca,Buildings}. Fibonacci-anyon synthesis uses Gauss complexity \cite{KliuchnikovEtAl}. Here denominators bound the minimum length of every evolving pair. For the specific braid studied by Kolev \cite[Section~5.1]{Kolev}, Theorem~\ref{thm:sharpwords} proves that its free words remain shortest after passing to the projective gate group. General existence results for faithful representations of free products \cite{MintyFreeProducts} do not identify this marked representation.

The character-variety dynamics of \cite{JangTropical,CantatLoray,CantatBers,SpaldingVeselov,Hadari} provide the setting for Theorem~\ref{thm:symmetricorbit}. Section~\ref{sec:fullorbit} compares the hypotheses of Jang's tropical lifting results and the specialized Farey trace recurrence \cite{FareyWords}. Maloni, Palesi and Tan obtain Fibonacci growth for $\log^+|\phi|$ under Bowditch conditions \cite[Theorem~4.8]{MPT}; top-degree trace terms also occur in plumbing coordinates \cite{MS}. Here the unit leading coefficients give exact non-Archimedean valuations even at initial points with bounded traces in the chosen embedding. Dubrovin and Mazzocco studied braid orbits of reflection triples in a Painlev\'e VI family \cite{DubrovinMazzocco}, and Iwasaki and Uehara counted periodic solutions via cubic-surface dynamics \cite{IwasakiUehara}. For finite orbits of reflection data see \cite{LamLandesmanLitt,Vayalinkal}. An infinite orbit of a pseudo-Anosov braid is also Zariski dense by Cantat's criterion: a suitable positive power acts as a hyperbolic element of his group and preserves no algebraic curve \cite[Corollary~3.4]{CantatBers}. Proposition~\ref{prop:zariski} proves density directly from valuations. A quantized character-variety counterpart to the braid action appears in \cite{BenZviBrochierJordan}.

The dense image in Theorem~\ref{thm:highrank} is the image of the free generators after convolution. Its three-dimensional invariant subspace for the first three generators identifies the evolving targets. The denominator estimate for every generator and inverse is what preserves the lower bound when all $r$ gates are available. Density on overlapping planes is related to the bridge lemma \cite{AA}; density for Jones braid representations is treated in \cite{FLW}. Long constructed braid group representations from free-group data \cite{LongBraid}. The relation of the Long--Moody construction to middle convolution is developed in \cite{HiroeNegami}, and related transformations of Painlev\'e character data appear in \cite{Boalch,DalMartello}.

\paragraph{Hypergeometric integrals and monodromy.}
Schechtman and Varchenko constructed solutions of the KZ equations as multidimensional hypergeometric integrals \cite{SV}. Continuing these solutions around a loop amounts to transporting their integration cycles. In another setting, Chestnov et al. derived Pfaffian systems for Feynman integrals from GKZ systems and used them to compute relations among the integrals \cite{Chestnov}. Can one exhibit a family of hypergeometric integrals and a basis of solutions in which continuation along explicit loops gives the reflection matrices constructed here?

Myers, Sati and Schreiber model KZ transport on conformal blocks in homotopy type theory \cite[Theorem~6.8]{MyersSatiSchreiber}. An explicit monodromy realization of our matrices would give concrete generators and braid identities to compare with that approach to formal verification.

\paragraph{Possible future applications.}
An application to acoustic data would require a choice of three complex input features, accounting for encoding and measurement costs, and a comparison with a matched classical baseline.

\section*{Acknowledgments}
This work was supported by the Research Institute for Mathematical Sciences, an International Joint Usage/Research Center located in Kyoto University. The ideas for this work grew out of the author's participation in the RIMS project ``Developments in Probabilistic and Variational Methods for Dynamical Systems.'' The author thanks the organizers and participants, as well as Mohammed Barakat, Keisuke Fujii, Tasuku Inui, Eiko Kin, Urs Schreiber, and Mao Shinoda for helpful discussions.

The author used ChatGPT (OpenAI) and Claude (Anthropic) for AI-assisted preparation of the manuscript.

\appendix
\section{Comparison with canonical-height theory}\label{app:height}
For a nonperiodic algebraic trace point, canonical heights give the asymptotic exponent independently of the local finite-time bounds.

\begin{definition}[Heights]\label{def:heights}
For homogeneous coordinates $[a_0:\cdots:a_m]$ over a number field $E$, the absolute logarithmic Weil height is
\[
 h([a_0:\cdots:a_m])=
 \frac1{[E:\Q]}\sum_v [E_v:\Q_v]\log\max_i|a_i|_v,
\]
where $|p|_v=p^{-1}$ at a place above $p$ and the usual modulus is used at archimedean places. The scalar height is $h(a)=h([1:a])$, and the affine matrix height is $h_{\mathrm{aff}}(W)=h([1:\text{entries of }W])$.
\end{definition}

\begin{prop}[Qualitative exponents from existing height theory]\label{rem:heightprior}
In this proposition, let $E$ be a number field containing the input data and $\lambda$.
Take the algebraic input data of Definition~\ref{def:ranktwoinput}, a pseudo-Anosov three-braid $\beta$, and $\lambda\in E^\times$ satisfying
\[
 \lambda\ne1,\qquad \lambda q\ne1,\qquad \det(\lambda P_{\rm in}-I_2)\ne0.
\]
Let $\boldsymbol M$ be the convolution realization in \eqref{eq:generalM}. If the initial normalized trace point is nonperiodic under $\beta$, then every pair $i\ne j$ has
\[
 \lim_{n\to\infty}L^{ij}_{\boldsymbol g,\beta}(n)^{1/n}
 =\lim_{n\to\infty}L^{ij}_{\boldsymbol M,\beta}(n)^{1/n}=\lambda_\beta.
\]
The conclusion holds without the local unit and invariant-region hypotheses.
\end{prop}
\begin{proof}
This is a consequence of \cite[Propositions~10.1 and 11.18, Corollary~11.25]{Abboud}. Let $f$ be the Fricke-surface automorphism induced by a pseudo-Anosov braid $\beta$. Choose $k\ge1$ so that $f^k$ belongs to the subgroup in \cite[Theorem~4.7 of the extended manuscript]{CantatLoray}. That theorem identifies $\lambda_1(f^k)$ with the spectral radius of its modular matrix, which is $\lambda_\beta^k$ by Proposition~\ref{prop:braidmatrices}. Thus $\lambda_1(f)=\lambda_\beta$.

We first check the surface hypotheses. The Fricke surfaces are geometrically integral hypersurfaces by the discriminant argument above. Their singular loci are finite: subtracting gradient equations gives $(x-y)(2-z)=(y-z)(2-x)=(z-x)(2-y)=0$. If all coordinates agree, the gradient equation is quadratic. Otherwise one coordinate is $2$, a second is also $2$, and the remaining one is fixed by the gradient equation. Being hypersurfaces, they are Cohen--Macaulay; regularity in codimension one therefore gives normality by Serre's criterion. Their projective closures are smooth at the infinity triangle, whose intersection matrix has diagonal $-1$ and off-diagonal $1$. A unit on the affine surface has principal divisor supported on these three boundary lines. Its intersection with each line is zero. The intersection matrix is invertible, so every coefficient of that divisor vanishes. Normality and projectivity then force the unit to be constant. Thus the surface is not $\mathbb G_m^2$, and \cite[Theorem~8.2]{Abboud} gives $\operatorname{QAlb}(S)=0$, where $\operatorname{QAlb}(S)$ is the quasi-Albanese variety of $S$ \cite[Section~8]{Abboud}.

For a nonconstant regular function $u:S\to\mathbb A^1$, resolve its extension to a morphism on a projective completion. By the cofinality assertion of \cite[Proposition~10.1]{Abboud}, choose a completion $X$ satisfying that proposition and dominating both the resolution and the original projective Fricke closure. Write $\bar u:X\to\mathbb P^1$ for the resulting morphism. The divisor $D=\bar u^*(\infty)$ is nonzero, effective, nef, and supported at infinity. Give it the pullback of the standard adelic metric on $\mathcal O_{\mathbb P^1}(1)$, so $h_{\bar D}=h\circ u$ on $S$. Write $\theta^-_X$ for the class on $X$ of Abboud's nef eigenclass for the inverse map. Its boundary coefficients are positive on this completion \cite[Proposition~10.1(6)--(7)]{Abboud}. Let $H$ be the pullback of the hyperplane at infinity from the original closure. A general fiber of $\bar u$ is linearly equivalent to $D$ and contains a curve meeting the affine surface. Its image has positive hyperplane degree, so the projection formula gives $D\cdot H>0$. As $H$ is effective and supported on the boundary and $D$ is nef, at least one boundary component meets $D$ positively and all meet it nonnegatively. Positivity of the boundary coefficients of $\theta^-_X$ gives $D\cdot\theta^-_X>0$.

Let $P$ be the nonperiodic initial trace point and write $\widehat h^+=h_{\overline{\theta^+}}$ for the forward canonical height in the notation of \cite[Section~11]{Abboud}. Corollary~11.25 there gives $\widehat h^+(P)>0$. Evaluating the adelic convergence of \cite[Proposition~11.18]{Abboud} at $P$ therefore gives
\[
 \lambda_\beta^{-n}h(u(f^nP))\longrightarrow
 (D\cdot\theta^-_X)\widehat h^+(P)>0.
\]
For a fixed algebraic matrix alphabet, the affine height $h_{\mathrm{aff}}(W)$ of a word of length $\ell$ is $O(\ell+1)$. This remains useful for projective representatives: if $W=cA_{j,n}$ in dimension $d$, then $c$ belongs to the coefficient field and
$d\,h(c)\le h(\det W)+h(\det A_{j,n})=O(\ell+1)$,
since Hurwitz moves preserve the finite list of component determinants. Consequently the height of a target pair trace is $O(\ell_i+\ell_j+1)$. Apply the preceding coordinate limit to the two affine trace expressions in Lemma~\ref{lem:transport}, then use Lemma~\ref{lem:upper}. This proves every pair exponent. The convolution identities used here are algebraic and require only the three displayed spectral exclusions, by Corollary~\ref{cor:mctraces} and Lemma~\ref{lem:transport}.
\end{proof}
The local proof of Proposition~\ref{thm:main} supplies the finite test for entry into the invariant region and the explicit inequalities \eqref{eq:transferfinite}, which retain information not present in the qualitative exponent statement.

\section{Local word-length bounds at a fixed rational point}\label{app:local}
Proposition~\ref{prop:f53} determined the first two pair lengths by reduction modulo $53$. At a place above five, the leading residues give an improved bound for every iterate, and show why the pole order together with a fixed number of normalized $5$-adic digits cannot determine every right-letter update.
\subsection{The local calculation}

Fix $N=13$, $c=3/5$, and use the local letters $a=M_1$, $b=M_2$, $h=M_3$ for the matrices in \eqref{eq:familyM}. The trace parameters from Section~\ref{sec:fullorbit} will not occur in this appendix. Put $q=e^{2\pi i/13}$ and $\tau=3(q-1)/5$.
Choose the place of $\mathbb Q(q)$ above $5$ for which the residue
$\alpha=\bar q$ has minimal polynomial
\[
f(X)=X^4+X^3-X^2+X+1\in\mathbb F_5[X].
\]
It is a factor of $\Phi_{13}$; every primitive thirteenth root over
$\mathbb F_5$ has degree $\operatorname{ord}_{13}(5)=4$. The completion
$L$ is unramified of degree four, with residue field $k=\mathbb F_{625}$.
Normalize $v(5)=1$, and for a word matrix $G$ put
\[
\pi_5(G)=-\min_{i,j}v(G_{ij}),\qquad
R_5(G)=\overline{5^{\pi_5(G)}G}\in M_3(k).
\]
All word determinants are units, so $\pi_5(G)\ge0$. Thus $\pi_5(G)$ is the least denominator exponent at this place: it is the least integer $e\ge0$ for which $5^eG$ has integral entries. Write $[R_5(G)]$ for
the class of this nonzero matrix modulo $k^\times$; it need not be invertible.
All six letters have pole order one.

The matrices in $H=\langle a,h\rangle$ have pole order at most one.
For example, ordering coordinates as $(e_1,e_3,e_2)$ gives blocks
$\left(\begin{smallmatrix}P&\tau u\\0&1\end{smallmatrix}\right)$
with $P\in GL_2(\mathcal O_L)$ and $u\in\mathcal O_L^2$.
For the blocks below, write
\[
R_5(g)=n_g e_2^T,\qquad
\ell=3(\alpha-1)(-e_1^T+e_3^T).
\]
Then $R_5(b)=e_2\ell$ and $R_5(b^{-1})=\alpha^{-1}e_2\ell$.

\subsection{A finite description of all target prefixes}
Write $F_0=0$, $F_1=1$, and $F_{j+2}=F_{j+1}+F_j$.
Define the words $(A_0,B_0,C_0)=(a,b,h)$ and
\[
(A_{n+1},B_{n+1},C_{n+1})=(A_nC_nA_n^{-1},A_n,C_n^{-1}B_nC_n).
\]
We consider prefixes of the words at times $n\ge1$, including the empty
prefix. Put $W_{\rm com}=aha^{-1}h^{-1}$ and
\[
\mathcal I=\{hah^{-1},ha^{-1}h^{-1},hah^{-1}a^{-1}h^{-1},haha^{-1}h^{-1}\}.
\]
An $H$-block is a maximal subword in $a^{\pm1},h^{\pm1}$. By Lemma~\ref{lem:freeproductgeodesics} and its proof, for $n\ge3$ the boundary blocks of $A_n,B_n$ are $W_{\rm com},W_{\rm com}^{-1}$ and those of $C_n$ are $h^{-1},h$. All internal blocks lie in $\mathcal I$, separated by single $b$-letters. The words at $n=1,2$ follow directly from the recurrence.
All nonempty $H$-block prefixes that occur belong to the following
15-element list. The last column records $\ell n_g$ as coefficients
in the basis $(1,\alpha,\alpha^2,\alpha^3)$; every entry is nonzero.
In each row $\pi_5(g)=1$, and every entry of $R_5(g)$ outside column two
vanishes. The table follows by multiplication in $k$.
\begin{center}
\begin{tabular}{lc@{\qquad}lc}
\toprule
$g$ & $\ell n_g$ & $g$ & $\ell n_g$\\
\midrule
$h^{-1}$ & $(2,2,4,4)$ & $ha^{-1}$ & $(2,1,2,0)$\\
$a$ & $(1,3,1,0)$ & $ha^{-1}h^{-1}$ & $(2,2,4,4)$\\
$ah$ & $(3,3,0,4)$ & $ha$ & $(1,4,4,1)$\\
$aha^{-1}$ & $(1,4,1,0)$ & $hah^{-1}$ & $(1,3,1,0)$\\
$W_{\rm com}$ & $(4,2,4,2)$ & $W_{\rm com}^{-1}$ & $(2,4,1,3)$\\
$h$ & $(1,3,1,0)$ & $W_{\rm com}^{-1}h^{-1}$ & $(2,2,4,4)$\\
$hah$ & $(2,0,2,4)$ & $haha^{-1}$ & $(3,1,0,2)$\\
$hW_{\rm com}$ & $(1,3,1,0)$ & &\\
\bottomrule
\end{tabular}
\end{center}

\begin{prop}[Exact pole orders and eight projective residue classes]\label{prop:fixedpoles}
Every nonempty prefix of a target word satisfies
\[
\pi_5(\text{prefix})=
\#\{\text{$H$-blocks in the prefix}\}+\#\{\text{$b^{\pm1}$-letters}\}.
\]
The projective leading residues of all target prefixes form a set of
exactly eight elements. For every $n\ge1$,
\begin{equation}\label{eq:fixedpoles}
\pi_5(A_n)=2F_{2n-1}-1,\qquad
\pi_5(C_n)=2F_{2n-2}+3.
\end{equation}
\end{prop}
\begin{proof}
For a prefix ending with an $H$-block and containing $m$ letters
$b^{\pm1}$, the highest-pole product is, up to a power of $\alpha^{-1}$,
\[
n_{g_0}(\ell n_{g_1})\cdots(\ell n_{g_m})e_2^T.
\]
It is nonzero by the table, even when the last block is incomplete.
Its pole order is $2m+1$. A prefix ending with a $b$-letter instead
has pole order $2m$ and leading matrix proportional to $n_{g_0}\ell$.
This proves the counting formula and excludes any pole decrease along
a target prefix.

Before the first $b$-letter, the possible words are
$a,ah,aha^{-1},W_{\rm com},h^{-1}$. After the first $b$-letter, the first
$H$-block is either $W_{\rm com}$ or $h^{-1}$. Thus the set of projective residue classes is
\[
\{[I_3]\}\ \cup\
\{[n_g e_2^T]:g=a,ah,aha^{-1},W_{\rm com},h^{-1}\}\ \cup\
\{[n_{W_{\rm com}}\ell],[n_{h^{-1}}\ell]\}.
\]
These eight matrices are pairwise nonproportional, as direct calculation
in $k$ verifies; they already occur by time two. Finally, the numbers
of $b$-letters in $A_n,C_n$ are $F_{2n-1}-1,F_{2n-2}+1$.
These formulas follow from the same substitution recurrence with initial
count vector $(0,1,0)$. Both full words end with an $H$-block, proving
\eqref{eq:fixedpoles}.
\end{proof}

\begin{corollary}[A two-letter improvement of the local lower bound]
Let $\ell_S$ be word length in the projective images of
$S=\{a^{\pm1},b^{\pm1},h^{\pm1}\}$, and set
$L^{13}(n)=\ell_S([A_n])+\ell_S([C_n])$. At $N=13,c=3/5$,
\[
2F_{2n}+2\le L^{13}(n)\le2F_{2n+2}\qquad(n\ge1).
\]
\end{corollary}
\begin{proof}
If a word $W$ represents $[A_n]$, then $W=\gamma A_n$ with $v(\gamma)=0$
by determinants. Hence $\pi_5(A_n)=\pi_5(W)\le|W|$. Apply the same
argument to $C_n$ and sum \eqref{eq:fixedpoles}. The upper bound is
Theorem~\ref{thm:fibonacci}.
\end{proof}
The lower-to-upper ratio still tends to $\Lambda^{-1}$,
$\Lambda=(3+\sqrt5)/2$. These eight projective classes describe the targets'
leading residues. They do not determine the full word length.

\subsection{Why the same state cannot handle all competing words}
For a one-digit example, let $X=ab^{-1}$ and $Y=a^{-1}b$. Both have pole two and the same leading matrix: the top terms of $a^{-1}$ and $b^{-1}$ are $\alpha^{-1}$ times those of $a$ and $b$, respectively. After right multiplication by $b$, the first becomes $a$ of pole one. The second is $a^{-1}b^2$ of pole two, since $b^2=I+(1-q)(b-I)$ and $1-\alpha\ne0$. Even the last letter can coincide: $a^2b$ and $ab^2$ have equal poles and leading matrices, while appending $b^{-1}$ gives poles one and two. The next result allows any fixed number of digits.

\begin{theorem}[Obstruction for every fixed digit depth]\label{thm:fixeddigits}
Let $\Gamma=\langle a,b,h\rangle$ as a linear matrix group. For any fixed
integer $r\ge1$, the state
\[
S_r(G)=\left(\pi_5(G),\ 5^{\pi_5(G)}G\bmod5^r\mathcal O_L\right)
\]
does not admit deterministic right-multiplication updates for all six
letters on all of $\Gamma$. Specifically, there are $X,Y\in\Gamma$
with $S_r(X)=S_r(Y)$ and a common finite continuation whose resulting
pole orders differ.
\end{theorem}
\begin{proof}
Let $T=ab$ and retain $W_{\rm com}=[a,h]$. The restriction of $T$ to
$E_{12}=\langle e_1,e_2\rangle$ is
\[
\begin{pmatrix}-q-\tau^2&-q\tau\\-\tau&-q\end{pmatrix}.
\]
Its trace has valuation $-2$ and determinant $q^2$ has valuation zero.
The two eigenvalues therefore have valuations $-2,2$; the third
eigenvalue of $T$ is $1$. Over a finite extension diagonalizing $T$,
let $\mathcal L_-$ be the eigenline of valuation $2$. It lies in $E_{12}$
and has both coordinates nonzero, since both off-diagonal entries
of the displayed block are nonzero.

The subgroup $H$ preserves $E_{13}=\langle e_1,e_3\rangle$ and fixes
\[
w=e_2+\frac{\tau}{q^2+1}(e_1-qe_3).
\]
The blocks in Proposition~\ref{prop:trianglesubgroup} place $H$ in
$\operatorname{diag}(GL_2(\mathcal O_L),1)$ in this basis, so its closure is compact.
Proposition~\ref{prop:arithmeticstructure} gives a Galois conjugate of
$W_{\rm com}|_{E_{13}}$ with trace greater than two and determinant one.
Its eigenvalues $\lambda,\lambda^{-1}$ are therefore not roots of unity. Every positive power of $W_{\rm com}$
has the same three distinct eigenlines: two in $E_{13}$ and $\langle w\rangle$.
None is $\mathcal L_-$: the latter has nonzero second coordinate and
zero third coordinate, while $w$ has nonzero third coordinate.

Compactness of the powers of $W_{\rm com}$ gives a positive integer $s$ such that
$P_s=W_{\rm com}^s\equiv I_3\pmod{5^r}$ in the original basis. More explicitly,
reduce the integral block modulo a sufficiently high power of $5$ and
take its finite order; a fixed extra precision compensates for the
change of basis. Then $P_s\in GL_3(\mathcal O_L)$, but $P_s$ does not
preserve $\mathcal L_-$. In a $T$-eigenbasis this gives a nonzero entry
in the $\mathcal L_-$ column outside its diagonal. Conjugation by $T^m$
multiplies that entry by an eigenvalue ratio of negative valuation.
Entry norms in two fixed bases differ by bounded multiplicative factors.
Therefore $\pi_5(T^mP_sT^{-m})\to\infty$ as $m\to\infty$ in the original basis too.

Set $X=T^m$ and $Y=T^mP_s$. Right multiplication by the integral
invertible matrix $P_s$ preserves the entry norm, so $\pi_5(X)=\pi_5(Y)$.
Moreover
\[
5^{\pi_5(X)}(Y-X)=5^{\pi_5(X)}X(P_s-I_3)\equiv0\pmod{5^r}.
\]
Hence $S_r(X)=S_r(Y)$. For sufficiently large $m$, the common right
continuation $T^{-m}$ gives $I_3$ from $X$, of pole zero, and
$T^mP_sT^{-m}$ from $Y$, of positive pole. If letter-by-letter updates
were determined by $S_r$, equal initial states would remain equal
through this continuation, a contradiction.
\end{proof}

These collisions give identical residue data for different matrices and hence no shortening relation. A normal form or an update that retains other information could still determine the all-time shortest lengths at this rational point.

The scripts in \path{anc/exact_checks/fixed_rational/} reproduce the local calculations. In particular, \path{padic_states.py} checks the residue transitions and \path{check_exact_blocks.py} verifies the block table.

\end{document}